\documentclass[twocolumn, final]{aastex701}
\usepackage[dvipsnames]{xcolor}
\usepackage{xcolor}
\usepackage{hyperref}
\usepackage{colortbl}
\definecolor{band}{RGB}{245,240,255} 
\usepackage{amsmath}  
\usepackage{empheq}     
\usepackage{natbib}

\hypersetup{
    linkcolor=blue!80!black,    
    citecolor=blue!50!black,    
    filecolor=magenta!70!black, 
    urlcolor=blue!50!black,     
    colorlinks=true,             
    pdfborder={0 0 0}            
}

\shorttitle{Galaxy Size-Mass-Wavelength Relation}
\shortauthors{Merchant et al.}
\received{}
\revised{}
\accepted{}
\submitjournal{ApJ}

\newcommand{\db}{\texttt{Dense Basis}}
\newcommand{\galfit}{\texttt{GALFIT}}
\newcommand{\eazy}{\texttt{EAzY}}

\begin{document}

\title{CANUCS/Technicolor Data Release 2: A Catalogue of Galaxy Structural Parameters in up to 29 HST+JWST bands and a Multi-Wavelength Exploration of the Galaxy Size--Mass Relation at $0.6 < z \leq 4$}

\author[0009-0000-5385-8674]{Maya Merchant}
\affiliation{Department of Physics and Astronomy, York University, 4700 Keele St., Toronto, Ontario, M3J 1P3, Canada}
\email{merchm@yorku.ca}  

\author[0000-0002-8530-9765]{Lamiya A. Mowla}
\email{lmowla@wellesley.edu}
\affiliation{Whitin Observatory, Department of Physics and Astronomy, Wellesley College, 106 Central Street, Wellesley, MA 02481, USA}
\affiliation{Center for Astronomy, Space Science, and Astrophysics, Independent University Bangladesh, Dhaka, Bangladesh}

\author[0000-0002-4872-2294]{Georgios E. Magdis}
\email{georgios.magdis@nbi.ku.dk}
\affiliation{Niels Bohr Institute, University of Copenhagen, Jagtvej 128, DK-2200 Copenhagen N, Denmark}
\affiliation{DTU-Space, Technical University of Denmark, Elektrovej 327, 2800, Kgs. Lyngby, Denmark}
\affiliation{Cosmic Dawn Center (DAWN), Jagtvej 128, DK-2200 Copenhagen N, Denmark}

\author[0000-0002-9330-9108]{Adam Muzzin}
\email{muzzin@yorku.ca}
\affiliation{Department of Physics and Astronomy, York University, 4700 Keele St., Toronto, Ontario, M3J 1P3, Canada}

\author[0000-0002-4201-7367]{Chris J. Willott}
\email{chris.willott@nrc.ca}
\affiliation{National Research Council of Canada, Herzberg Astronomy \& Astrophysics Research Centre, 5071 West Saanich Road, Victoria, BC, V9E 2E7, Canada}

\author[0000-0002-4542-921X]{Roberto Abraham}
\email{roberto.abraham@utoronto.ca}
\affiliation{David A. Dunlap Department of Astronomy and Astrophysics, University of Toronto, 50 St. George Street, Toronto, Ontario, M5S 3H4, Canada}
\affiliation{Dunlap Institute for Astronomy and Astrophysics, 50 St. George Street, Toronto, Ontario, M5S 3H4, Canada}

\author[0000-0003-3983-5438]{Yoshihisa Asada}
\email{yoshi.asada@utoronto.ca}
\affiliation{Dunlap Institute for Astronomy and Astrophysics, 50 St. George Street, Toronto, Ontario, M5S 3H4, Canada}

\author[0000-0001-5984-0395]{Maru\v{s}a Brada\v{c}}
\email{marusa.bradac@fmf.uni-lj.si}
\affiliation{Faculty of Mathematics and Physics, Jadranska ulica 19, SI-1000 Ljubljana, Slovenia}
\affiliation{Department of Physics and Astronomy, University of California Davis, 1 Shields Avenue, Davis, CA 95616, USA}

\author[0000-0003-2680-005X]{Gabriel B. Brammer}
\email{gabriel.brammer@nbi.ku.dk}
\affiliation{Niels Bohr Institute, University of Copenhagen, Jagtvej 128, DK-2200 Copenhagen N, Denmark}
\affiliation{Cosmic Dawn Center (DAWN), Jagtvej 128, DK-2200 Copenhagen N, Denmark}

\author[0000-0001-8325-1742]{Guillaume Desprez}
\email{guillaume.desprez@protonmail.com}
\affiliation{Kapteyn Astronomical Institute, University of Groningen, P.O. Box 800, 9700AV Groningen, The Netherlands}

\author[0000-0001-9298-3523]{Kartheik G. Iyer}
\email{kgi2103@columbia.edu}
\affiliation{Columbia Astrophysics Laboratory, Columbia University, 550 West 120th Street, New York, NY 10027, USA}

\author[0000-0003-3243-9969]{Nicholas S. Martis}
\email{nicholas.martis@fmf.uni-lj.si}
\affiliation{Faculty of Mathematics and Physics, Jadranska ulica 19, SI-1000 Ljubljana, Slovenia}

\author{Ga\"el Noirot}
\email{gnoirot@stsci.edu}
\affiliation{Space Telescope Science Institute, 3700 San Martin Drive, Baltimore, Maryland 21218, USA}

\author[0009-0009-4388-898X]{Gregor Rihtar\v{s}i\v{c}}
\email{gregor.rihtarsic@fmf.uni-lj.si}
\affiliation{Faculty of Mathematics and Physics, Jadranska ulica 19, SI-1000 Ljubljana, Slovenia}

\author[0000-0002-7712-7857]{Marcin Sawicki}
\email{marcin.sawicki@smu.ca}
\affiliation{Department of Astronomy and Physics and Institute for Computational Astrophysics, Saint Mary's University, 923 Robie Street, Halifax, Nova Scotia B3H 3C3, Canada}

\author[0000-0001-8830-2166]{Ghassan T. E. Sarrouh}
\email{gsarrouh@yorku.ca}
\affiliation{Department of Physics and Astronomy, York University, 4700 Keele St., Toronto, Ontario, M3J 1P3, Canada}

\author[0009-0000-8716-7695]{Sunna Withers}
\email{sunnaw@yorku.ca}
\affiliation{Department of Physics and Astronomy, York University, 4700 Keele St., Toronto, Ontario, M3J 1P3, Canada}

\author[0000-0001-9610-7950]{Natalie Allen}
\email{natalie.allen@nbi.ku.dk}
\affiliation{Niels Bohr Institute, University of Copenhagen, Jagtvej 128, DK-2200 Copenhagen N, Denmark}

\author[0000-0002-0243-6575]{Jacqueline Antwi-Danso}
\email{j.antwidanso@utoronto.ca}
\affiliation{David A. Dunlap Department of Astronomy and Astrophysics, University of Toronto, 50 St. George Street, Toronto, Ontario, M5S 3H4, Canada}

\author[0000-0002-6741-078X]{Westley Brown}
\email{westleyb@yorku.ca}
\affiliation{Department of Physics and Astronomy, York University, 4700 Keele St., Toronto, Ontario, M3J 1P3, Canada}

\author[0009-0000-2101-1938]{Jon Jude\v{z}}
\email{jon.judez@fmf.uni-lj.si}
\affiliation{Faculty of Mathematics and Physics, Jadranska ulica 19, SI-1000 Ljubljana, Slovenia}

\author[0000-0001-9002-3502]{Danilo Marchesini}
\email{danilo.marchesini@tufts.edu}
\affiliation{Department of Physics and Astronomy, Tufts University, 574 Boston Avenue, Suite 304, Medford, MA 02155, USA}

\author[0000-0001-8115-5845]{Rosa M. M\'erida}
\email{rosa.meridagonzalez@smu.ca}
\affiliation{Department of Astronomy and Physics and Institute for Computational Astrophysics, Saint Mary's University, 923 Robie Street, Halifax, Nova Scotia B3H 3C3, Canada}

\author[0009-0009-2307-2350]{Katherine Myers}
\email{kjmyers@yorku.ca}
\affiliation{Department of Physics and Astronomy, York University, 4700 Keele St., Toronto, Ontario, M3J 1P3, Canada}

\author[0000-0002-6265-2675]{Luke Robbins}
\email{andrew.robbins@tufts.edu}
\affiliation{Department of Physics and Astronomy, Tufts University, 574 Boston Avenue, Suite 304, Medford, MA 02155, USA}

\author[0000-0003-0780-9526]{Visal Sok}
\email{visal.sok@colorado.edu}
\affiliation{Department of Astrophysical and Planetary Sciences, University of Colorado, 2000 Colorado Ave, Boulder, CO 80309, USA}

\begin{abstract}
We present James Webb Space Telescope (JWST) results of a morphological study of galaxies in the CAnadian NIRISS Unbiased Cluster (CANUCS) and Technicolor surveys, observed in 19 medium- and broadband NIRCam filters in five CANUCS NIRCam Flanking Fields with rest-frame wavelength coverage between $\sim 0.2 - 3.2\mu m$. Using \galfit\ , we measure the morphological parameters of $\sim$ 4,100 star-forming galaxies at $0.6 < z \leq 4$ with stellar masses of $8.5 < \text{log}(M_*/M_\odot) \leq 11.5$. This enables us to concurrently examine how galaxy size varies as a function of stellar mass, redshift, and rest-frame wavelength to provide a novel parametrization of the galaxy size--wavelength relation. Additionally, we analyze the evolution of the galaxy size--mass relation in the rest-frame optical and NIR with the introduction of wavelength as a free parameter. We report a gradient in the slope of the size--mass relation with respect to rest-frame wavelength with a critical crossover mass at $\sim 10^{9.5} M_\odot$. We propose this characteristic mass as the stellar mass at which galaxies transition between diffuse and compact morphologies. We concurrently present the data release\footnote{\url{https://niriss.github.io/data.html}} of morphological measurements of the five CANUCS-Technicolor NIRCam Flanking Fields in which we provide structural parameters for $\sim$ 41,000 galaxies in up to 29 JWST+HST filters.
\end{abstract}


\section{Introduction} 
Galaxy size has long been used to trace the physical drivers of galaxy evolution and to gain insight into galaxy formation histories (e.g. \citealt{zhang_intro, Conselice_2014, zhang2026bandiglobalmorphology, miller2026bandiirelationshipoptical}), stellar populations (e.g. \citealt{Conroy_2013}), mass assembly (e.g. \citealt{Naab_2009}), dark matter halos (e.g. \citealt{Mo_1998, Kravtsov_2013, Jiang_2019}) and stellar disks (e.g. \citealt{vanDokkum_2013}). Comparisons between galaxy size and physical properties such as stellar mass and star formation history have proven valuable for distinguishing different modes of galaxy evolution (e.g. \citealt{Courteau_2014, mimi_song_sfh, Yang_2019, mcleod_sfh}) as various growth mechanisms (i.e. mergers and star formation) having distinct impacts on morphological development. Consequently, the distribution of galaxy sizes at different stellar masses provides critical insight into their formation channels \citep{Ribeiro_2016}.

\begin{figure*}[htp!]
    \centering
    \includegraphics[width=\linewidth]{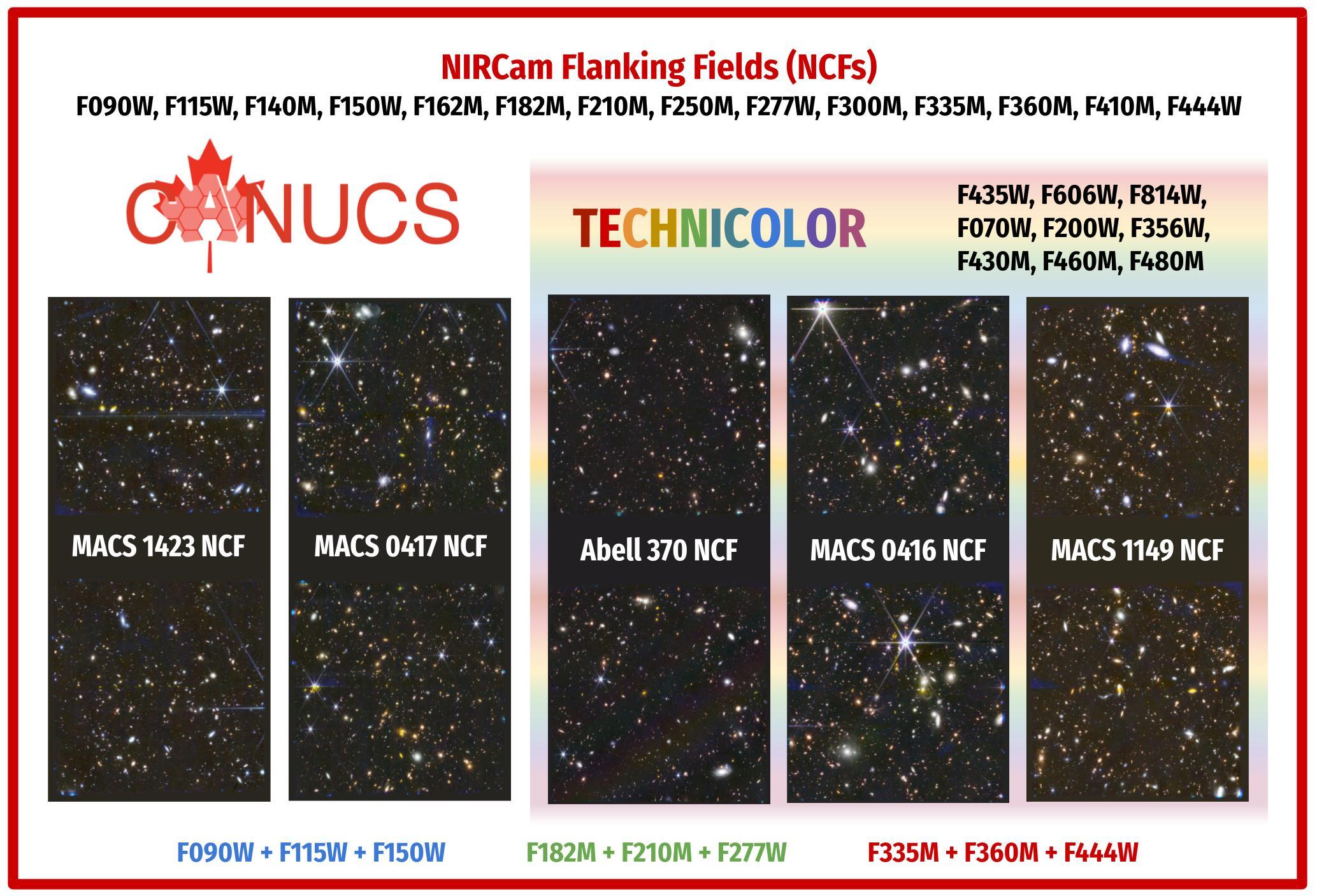}
    \caption{RGB mosaics of the five CANUCS NIRCam Flanking Fields (NCF): MACS J1423, MACS J0417, Abell 370, MACS J0416, and MACS J1149. The RGB images are created from the combination of F090W + F115W + F150W (blue), F182M + F210M + F277W (green), and F335M + F360M + F444W (red) filters. Each pair of squares represents the two NIRCam modules, each covering approximately $2.2' \times 2.2'$ on the sky. The CANUCS survey provides coverage in fourteen wide- and medium- bands across all five fields (excluding F162M and F250M from MACS J1149), while the Technicolor survey provides coverage in Abell 370, MACS J0416, and MACS J1149 with an additional six wide- and medium-band filters and ancillary HST data.}
    \label{fig:rgbs}
\end{figure*}

The galaxy size--mass relation correlates these two physical parameters, and has been extensively studied in the local universe using ground-based surveys (e.g. \citealt{lange-2015, Kawinwanichakij_2021, George2024, George2025}) and traced up to Cosmic Noon through large Hubble Space Telescope (HST) extragalactic programs (e.g. \citealt{vanderWel2012_snr, vanderWel_2014, Mowla_2019}). Studies of the size--mass relation  consistently show that galaxy size increases as a function of stellar mass (e.g. \citealt{Ferguson_2004, Trujillo_2006, Mosleh_2012, lange-2015, miller2024jwstuncoversopticalsize, miller2026bandiirelationshipoptical}). At a fixed stellar mass, star-forming galaxies tend to be larger than quiescent galaxies, with prominent disk structures while quiescent galaxies are generally more compact and spheroidal (e.g. \citealt{williams_sizemass, ilbert_sizemass, Straatman_sizemass, ito_sizemass, George2024}) which has generally been observed up to $z \sim 5.5$ \citep{Ward_2024}.

However, many measurements of the size--mass relation have relied on single-band imaging, causing the probed rest-frame wavelengths to vary across redshift. To enable comparisons across wavelengths, empirical colour-gradient corrections are often applied to estimate galaxy sizes at a fixed rest-frame wavelength, most commonly 0.5~$\mu$m (e.g. \citealt{vanderWel2012_snr}). This approach introduces systematic uncertainties as light is a biased tracer: it depends on stellar population age, metallicity, and the geometry of dust attenuation. Incorporating wavelength dependence is essential to our understanding of galaxy evolution and is therefore required to accurately trace the evolution of galaxy size (e.g. \citealt{Suess_2019, miller2024jwstuncoversopticalsize, miller2026bandiirelationshipoptical, George2024, George2025}). 

Furthermore, past approaches to the size--mass relation often relied on simplifying or nonphysical assumptions, such as adopting uniform stellar populations or consistent dust models across all systems. Such assumptions can bias inferred structural parameters and obscure the true physical diversity of galaxy morphologies. Star-forming galaxies are particularly prone to these effects due to their complex structures and diverse stellar populations \citep{shapley2023jwstnirspecbalmerlinemeasurementsstar, G_mez_Guijarro_2023, Gillman_2024}. They contain stars of different ages, clumpy regions of star formation, disky morphologies affected by inclination, and varying degrees of dust obscuration \citep{Sok2025, Tan2025}. To study the evolution of star-forming galaxy morphology, it is therefore essential to construct a rest-frame wavelength-dependent size--mass relation to account for variations in stellar population age, dust geometry, and star formation distribution.

Previous HST studies have extensively investigated structural properties using rest-frame optical imaging with HST's Wide Field Camera 3 (WFC3) at $z \lesssim 3$ (e.g. \citealt{vanderWel_2014, Morishita_2014, Huertas_Company_2016}); however, HST faced significant limitations in wavelength coverage beyond the rest-optical, where the limits of WFC3 are surpassed \citep{Treu2023}. Thus, HST morphological studies at $z > 3$ (e.g. \citealt{Conselice2009, Shibuya2015, Ribeiro_2016}) relied on rest-frame UV measurements. In this regime, emission is dominated by young stellar populations and thus inferred galaxy morphology is traced from active star formation, rather than the full underlying stellar mass distribution \citep{Treu2023}. The necessity of multi-wavelength observations became increasingly evident following the launch of JWST, which enabled morphological study with the remarkable ability to resolve galaxy morphologies down to $\lesssim 100$ pc across the rest-frame optical and UV (e.g. \citealt{Morishita2024, Finkelstein2022, Naidu2022, Treu2023}).

Radial colour gradients serve as a key framework for understanding multi-band trends of physical size; they describe the gradient in the mass-to-light ratio \citep{Bell_ML_ratio} as driven by dust attenuation, metallicity, or stellar age, compared between central and outskirt regions. Previous studies of the size--mass relation (e.g. \citealt{Morishita_2014, Suess_2019, Treu2023, Allen2024, George2024, George2025}) attest that evolution of the size-mass relation is primarily due to colour gradients, indicating the stellar mass build-up, stellar population, and age of the galaxy varies with distance from the galactic center. 

The capabilities of JWST now allow direct measurements from rest-frame ultraviolet (UV) to near-infrared (NIR) light for galaxies at Cosmic Noon, revealing striking wavelength-dependent differences in morphology and providing new insights into how galaxies assemble their stellar mass and structure over cosmic time (e.g. \citealt{George2025, chen2026bendingsizemassrelationstarforming, miller2026bandiirelationshipoptical, lilian-Yang2025, McGrath_2026}). 

With rest-frame wavelength coverage from JWST/NIRCam between $\sim 0.2 - 3.2 \mu$m up to $z=4$, we seek a framework to describe the relationship between physical size, redshift, stellar mass, and rest-frame wavelength concurrently at $0.6 < z \leq 4$. We expand on the existing size-mass relation with a multi-wavelength analysis to test the effects that wavelength has on the size--mass slope, and interpret physical drivers of these trends. 

The paper is organized as follows: Section \ref{sec:data products} introduces the data products used as well as the SED-fitting processes from which key physical parameters such as redshift and stellar mass are extracted. Section \ref{sec:morph modeling} describes the size determination method and sample selection procedures used, and introduces data products included in the catalogue release. Section \ref{results} presents the results of the rest-optical and rest-NIR size--mass relation in comparison to literature, as well as an analysis of the size--wavelength relation in regards to effects of redshift and stellar mass. Section \ref{discussion} extends this work by analyzing the multi-wavelength galaxy size--mass relation and investigating the trends that arise therein. A summary is given in Section \ref{sec:summary}.

We adopt a flat $\Lambda$CDM cosmology with $H_0$ = 70 km s$^{-1}$ Mpc$^{-1}$, and $\Omega_{M,0}$ = 0.3. We assume a Chabrier initial mass function \citep{Chabrier_2003} and a Calzetti dust attenuation law \citep{Calzetti_2000}. Magnitudes are expressed in the AB system \citep{Oke-gunn1983}.

\section{Data and Measurements} \label{sec:data products}
This study uses data from the JWST Canadian NIRISS Unbiased Cluster Survey (CANUCS; PI: Willott), which obtained NIRCam and JWST/Near Infrared Imager and Slitless Spectrograph (NIRISS) imaging of five massive galaxy clusters: Abell~370, MACS~J0416.1$-$2403, MACS~J0417.5$-$1154, MACS~J1149.5+2223, and MACS~J1423.8+2404, along with five associated NIRCam flanking fields (NCFs; \citealt{Willott_2022, Sarrouh2025}). In this work, we focus on the NCFs of these five lensing clusters (Figure \ref{fig:rgbs}) due to the availability of medium band filters for these fields and to preferentially avoid uncertainties from shear, magnification, and contamination from intracluster light. While magnification is considerably reduced in the NCFs, we note it is not fully negligible, with $\mu_{max} = 1.6$ in MACS J0417 and $\mu_{max} = 1.2$ in the remaining NCFs, where $\mu = 1$ represents no magnification (see \citealt{Sarrouh2025}). 

\subsection{Images}
\label{sec:images}
In Cycle 1, the NCFs were observed with a combination of wide- and medium-band NIRCam filters. The available wide-band imaging includes F090W, F115W, F150W, F277W, and F444W, while the medium-band coverage includes F140M, F162M, F182M, F210M, F250M, F300M, F335M, F360M, and F410M. Most filters have exposure times of approximately 10.3 ks, with shorter exposures of approximately 5.7 ks for F140M, F162M, F250M, and F300M. In the MACS~J1149 NCF, a program definition error resulted in the absence of F162M and F250M observations, with deeper exposures obtained instead in F150W and F277W. The CANUCS NCF dataset is further augmented by the Cycle 2 JWST in Technicolor program (ID 3362; PI: Muzzin), which provides additional wide-, medium-, and narrow-band NIRCam imaging for three NCFs: Abell 370, MACS J0416, and MACS J1149. This program adds imaging in the F070W, F164N, F187N, F200W, F356W, F430M, F460M, and F480M filters, with typical exposure times of approximately 10 ks with $3\sigma$ depths reaching $\sim29.5-30$ mag \citep{Sarrouh2025}. Combined with the Cycle 1 observations, these programs provide extensive NIRCam coverage from 0.7 to 4.4 $\mu$m across these three NCFs over an area of approximately 30 arcmin$^{2}$ \citep{Sarrouh2025}. The remaining two fields, MACS J0417 and MACS J1423, are covered by the Cycle 1 filter set only.

\begin{figure*}
    \centering
    \includegraphics[width=0.96\linewidth]{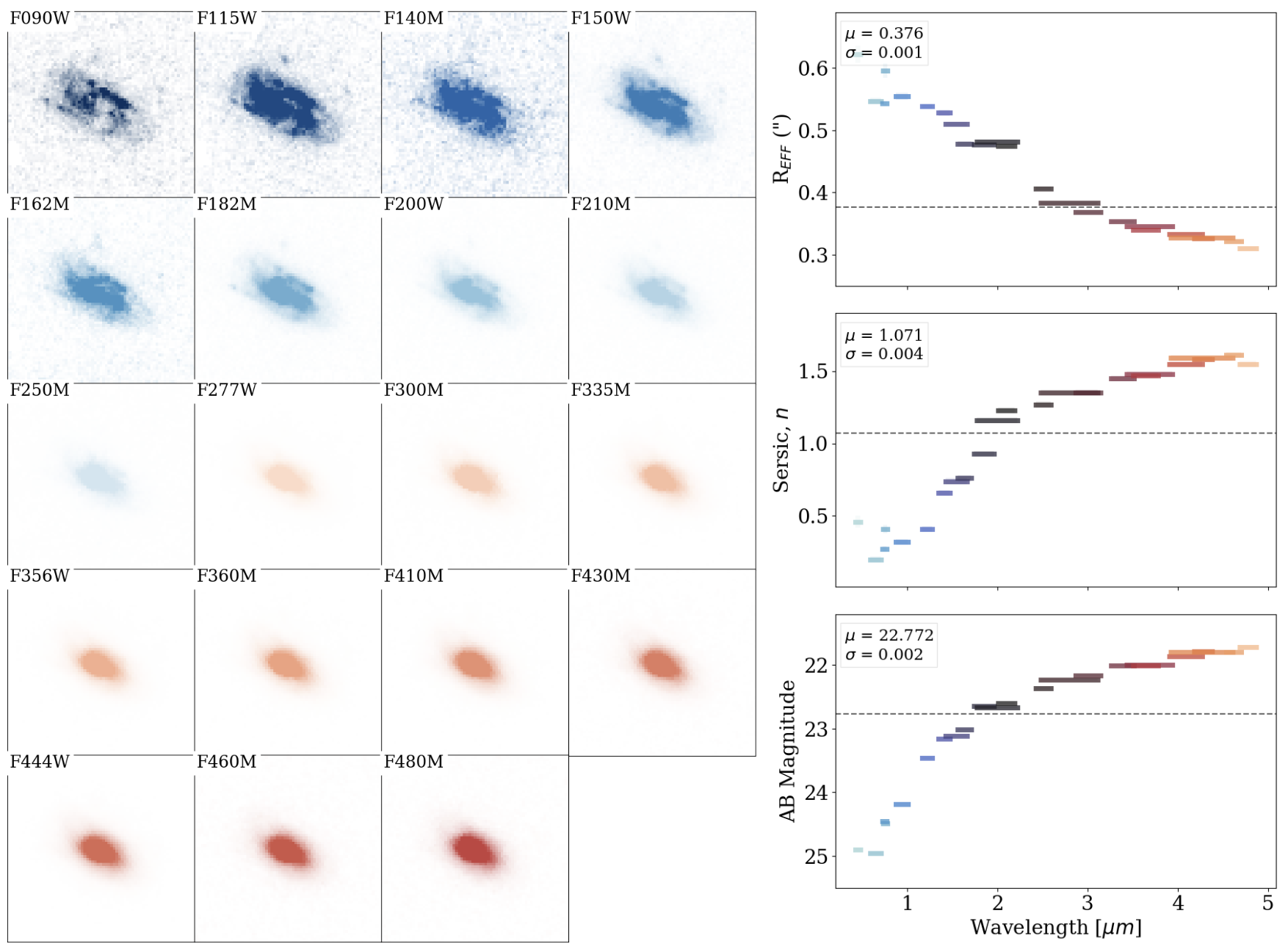}
    \caption{Left: Images of a single galaxy (CANUCS-A370-2200379; $\log(M_*/M_\odot) = 10.46 \pm 0.001$, $z  = 1.66 \pm 0.048$) across 19 NIRCam filters shown by their waverange. Right: Associated parameters of effective radius, Sérsic index, and magnitude as a function of observed wavelength for this galaxy. The mean ($\mu$; dashed horizontal line) and standard error ($\sigma$) is included for each variable.}
    \label{fig:19 filters}
\end{figure*}

The CANUCS fields are supplemented with ancillary HST imaging from multiple legacy programs. Abell~370, MACS~J0416, and MACS~J1149 were observed as part of the Hubble Frontier Fields program (HFF; PI: Lotz), which provides imaging in both the prime and parallel pointings, roughly corresponding to the CLU and NCF regions, respectively \citep{Lotz_2017_HFF}. For these three flanking fields, seven HST bands are available (F435W, F606W, F814W, F105W, F125W, F140W, and F160W) resulting in a total of 29 wide, medium, and narrow band filters. MACS~J0416, MACS~J1149, and MACSJ~J1423 were also observed as part of the Cluster Lensing And Supernova survey with Hubble (CLASH; PI: Postman; \citealt{Postman2012}). MACS~J0417 and MACS~J1423 are further supplemented with deep HST/WFC3 UVIS imaging in the F438W and F606W filters from program GO-16667 (PI: M.~Bradač).

Imaging data from \textit{HST} and \textit{JWST} are processed using the \texttt{grizli} software package \citep{grizli_brammer} and follow reduction procedures that are well documented in the literature (e.g.  \citealt{kokorev_hubble_2022}). Background subtraction is applied to the flanking-field images, which also mitigates residual image artifacts, including persistence. The RMS map is constructed by estimating the noise level in each pixel for a given filter and rescaling the corresponding weight image such that the sky background of the RMS-normalized science image follows a standard normal distribution. A comprehensive description of the CANUCS observing strategy, data reduction, and catalogue construction is presented in \cite{Sarrouh2025}.

While all available imaging is used for photometry and stellar population synthesis modelling, we restrict the morphology analysis to JWST wide- and medium-band images to take advantage of their higher signal-to-noise ratios and spatial resolution (see \citealt{Sarrouh2025}). From the 29 combined HST and JWST filters available, we exclude HST wide-band filters due to a larger point-spread function (PSF) and lower signal-to-noise resulting in skewed morphology measurements. Removing these filters improves the robustness of the size--wavelength fits without loss of relevant information. In addition, we exclude the shortest-wavelength filter NIRCam/F070W in this analysis as it is undersampled with a pixel scale comparable to the PSF at $\lambda_{\rm eff} \sim 0.7\,\mu{\rm m}$ which limits the effective resolution. The final morphology sample therefore consists of nineteen JWST filters, shown in Figure \ref{fig:19 filters} for a single galaxy. This figure shows the science images for CANUCS-A370-2200379 and the corresponding wavelength-dependent structural parameters (effective radius, Sérsic index, axis ratio, and AB magnitude) with wavelength uncertainties defined by the filter widths \citep{Rodrigo2012}. The process for deriving these structural parameters is discussed in depth in Section \ref{sec:morph modeling}. This image shows stark evidence of the drastic variation in structural parameters for a single galaxy as a function of observed wavelength, motivating the incorporation of wavelength as a free parameter in morphological analysis. For this analysis, we adopt the effective wavelength of each filter, which accounts for the full transmission curve and spectral shape. We define the wavelength uncertainty as the effective width, corresponding to the width of a top-hat filter with 100\% transmission that encloses the same total flux as the real filter.

\subsection{Photometry}
Source detection in CANUCS is performed on a deep $\chi_{\mathrm{mean}}$ detection image constructed by co-adding all available background-subtracted  JWST and optical HST images following a modified chi-squared technique \citep{Szalay1999, DrlicaWagner2018}, which enhances the detectability of faint sources while maintaining uniform noise properties across the field. HST/WFC3 IR images are excluded due to their broader point-spread function and larger native pixel scale. Individual filter images are inspected for artifacts, with masks applied during the co-addition stage to prevent spurious detections while preserving sensitivity in affected regions. Source detection and segmentation are carried out using the \texttt{photutils} \citep{bradley_photutils} implementation of \texttt{SourceExtractor} \citep{bertin_sextractor}, with a two-stage ``cold+hot'' strategy optimized separately for bright, extended sources and faint, compact sources. The two detection catalogues are merged following a GALAPAGOS-style approach \citep{barden_galapagos}, with field-dependent parameters. Bright stars and residual artifacts are excluded through manual masking, and detection thresholds are calibrated using the measured RMS of the $\chi_{\mathrm{mean}}$ image.

\subsection{SED Fitting}\label{subsec:sed fitting}
We utilize redshifts and stellar population parameters from the CANUCS catalogue as described in \citet{Sarrouh2025}. Photometric redshifts are computed using the template-fitting code \texttt{EAzY-py} \citep{Brammer2008-eazy}, derived using 0\farcs3-diameter aperture photometry in all available HST+JWST filters, with a 5\% systematic uncertainty added in quadrature. A magnitude prior and the IGM/CGM attenuation prescription of \citet{Asada2025} are applied; a low photometric outlier fraction of $\sim 4\%$ is noted in the flanking fields \citep{Sarrouh2025}. Stellar population properties are derived using the SED-fitting code \db\ \citep{Iyer_2017, Iyer_2019} which employs a non-parametric star formation history parameterized by mass-assembly quantiles, using FSPS-based stellar and nebular models \citep{conroy2010}. A Calzetti dust attenuation law \citep{Calzetti_2000} is adopted. In this paper, we use stellar population parameters from \db\ unless otherwise stated; in particular, we utilize the 50th percentile of the present stellar mass ($M_*/M_\odot$) of each galaxy and the star-formation rate (SFR; $M_\odot/yr$) averaged over the last 100 Myr \citep{Iyer_2019}.

\subsection{PSF Construction}
Accurate morphological modelling of galaxies requires an accurate PSF for each filter, as analytic surface-brightness models must be convolved with the instrumental PSF prior to fitting. Empirical PSFs are constructed by \citet{Sarrouh2025}, using carefully selected isolated point sources in each field. Stars are identified via their locus in surface-brightness versus magnitude space, visually inspected to reject saturated or contaminated sources, re-centered, and flux-normalized prior to stacking. For filters common to both cluster and flanking field pointings, a single PSF is constructed by combining stars from both fields, since these observations were obtained at identical position angles and closely spaced in time. For Technicolor filters observed in Cycle~2, PSF consistency across epochs is verified before combining stars with those from Cycle~1. The final PSF for each filter is constructed by merging the empirical PSF core with the outer regions of a scaled \texttt{STPSF} model \citep{perrin_webbpsf}, preserving the encircled energy while improving the signal-to-noise ratio at large radii. The resulting PSF models have an initial cutout size of $4\farcs04 \times 4\farcs04$ (101$\times$101 pixels), substantially larger than the typical galaxy size. To reduce computational cost during morphological fitting, we adopt a smaller PSF cutout of $0\farcs84 \times 0\farcs84$ (21$\times$21 pixels). We verify that this reduced PSF size does not bias the fitted structural parameters while significantly accelerating the fitting process.

\section{Galaxy Size Determination}\label{sec:morph modeling}
The broad wavelength coverage of the CANUCS-Technicolor imaging enables morphological measurements across a wide range of rest-frame optical and near-infrared wavelengths. We characterize galaxy structure by fitting single-component Sérsic profiles using \texttt{GALFIT} \citep{Peng_2002, Peng_2010}, and adopt the effective radius $R_{\mathrm{eff}}$, defined as the semi-major axis of the ellipse enclosing half the total flux of the best-fitting model, as our primary size measurement. The fitting procedure and sample selection are described in Sections~\ref{subsec:galfit} and~\ref{sec:data selection}, respectively.


\subsection{Size Measurements}
\label{subsec:galfit}
Galaxy images are prepared for structural fitting using the 40~mas pixel-scale CANUCS-Technicolor mosaics in all filters described in Section~\ref{sec:images} and size determination is performed following the method of \citet{Mowla_2019}. For each object, the cutout size is set to 15$\times$ the semi-major axis length $A$, derived from \texttt{photutils}, with a minimum cutout size of $2\farcs2$ to ensure sufficient sky background estimation for compact sources. Square cutouts are extracted from the background-subtracted science images and the corresponding RMS maps for each filter, along with the combined segmentation map. A single mask is constructed from the segmentation map for each object, in which all neighbouring sources are masked and only the target galaxy is left unmasked, preventing contamination from overlapping sources from biasing the structural parameters of the target; segmentation maps for each object are reported by \citet{Sarrouh2025}.

We fit single-component Sérsic profiles \citep{1968sersic} to each galaxy in each filter independently using \texttt{GALFIT} \citep{Peng_2002, Peng_2010}, supplying the image cutout, RMS map, PSF model, and segmentation mask. The free parameters in the fit are the total magnitude, the effective radius $R_\mathrm{eff}$ measured along the semi-major axis, the Sérsic index $n$, the axis ratio $Q$, the position angle, the central position $(x_0,~y_0)$, and an additive sky background term, with initial estimates drawn from the photometric catalogue. We impose physically motivated parameter constraints, requiring $0.1 < n < 10$, $0\farcs04 < R_\mathrm{eff} < 4\farcs0$ (corresponding to $0.1$--$100$ pixels at the 40~mas pixel scale), and $0.05 < Q < 1$, and allow the total magnitude to vary within $\pm3$~mag of the input \texttt{photutils} \texttt{MAG\_AUTO} value. These bounds are chosen to be sufficiently broad to accommodate the full range of galaxy morphologies present in the sample while excluding unphysical solutions that arise from fitting failures or low signal-to-noise data. Each galaxy is fit independently in each filter, allowing the structural parameters to vary freely with observed wavelength and enabling a direct measurement of the size--wavelength relation across the full CANUCS-Technicolor filter set.

We define a morphological uncertainty parameter on each \galfit\ measurement as $R_\mathrm{eff} / \sigma_{R_\mathrm{eff}}$, where $\sigma_{R_\mathrm{eff}}$ is the uncertainty on the effective radius returned by \galfit. This quantity serves as an uncertainty of the best-fit size measurement in a given filter and is used as a per-galaxy, per-filter weight in the subsequent size--wavelength analysis (Section~\ref{sec:size_wav}).

\subsection{CANUCS-Technicolor NCF Morphology Catalogue}
We present the CANUCS-Technicolor NCF Morphology Catalogue\footnote[1]{\url{https://niriss.github.io/data.html}}, which contains single-component Sérsic fit parameters measured with \galfit\ for galaxies in the five CANUCS NIRCam flanking fields. The parent sample is drawn from the CANUCS-Technicolor photometric catalogue \citep{Sarrouh2025}, from which non-point-source objects are selected with positive flux measured in a $0\farcs3$-diameter aperture in F444W. This criterion is designed to avoid biasing against galaxies that may be undetected in individual filters, including highly dust-obscured systems and sources with strong emission lines.

Applying these criteria yields a parent sample of 41,202 galaxies across the five flanking fields: 7,511 in MACS~J0416, 8,500 in MACS~J0417, 8,575 in MACS~J1149, 8,277 in MACS~J1423, and 8,339 in Abell~370. Of these, three fields (Abell~370, MACS~J0416, and MACS~J1149) benefit from the full CANUCS-Technicolor filter set and are covered in 23 filters with the exception of MACS~J1149, whose program definition error resulted in the loss of F162M and F250M for a total coverage in 21 filters. MACS~J0417 and MACS~J1423 are covered by the Cycle~1 filter set only and are thus covered in 14 and 16 filters respectively. A full description of HST+JWST coverage in each field is outlined in \citet{Sarrouh2025}.

To flag potentially unreliable measurements, we identify size measurements in individual filters for which the best-fit Sérsic index or axis ratio values do not converge within the \galfit\ parameter bounds ($0.1 < n < 10$ and $0.05 < Q < 1$). Each catalogue entry includes a \texttt{USE\_FLAG} indicating whether the fitted parameters satisfy these bounds (1) or not (0) in at least 10 filters. Approximately 34\% of individual filter measurements are flagged in this way (with $\sim 18\%$ of objects flagged for $n$ and $\sim 26\%$ for $Q$); users are advised to treat flagged measurements with caution. The flagging rate varies across filters, significantly higher in HST bands (F435W, F606W, F814W, F125W, F140M, and F160W) compared to JWST/NIRCam. In the NIRCam bands, higher flagging rates are displayed in shallower filters with a roughly 80\% increase in the number of affected galaxies between the deepest and shallowest filters (F277W and F480M respectively). Individual flags for the Sérsic index and axis ratio parameters (\texttt{USE\_N} and \texttt{USE\_Q}) are also included. 

A quality flag for position angle, $PA$, is also included (\texttt{USE\_PA}), although not incorporated in the combined \texttt{USE\_FLAG}. For each filter, a position angle measurement is considered unreliable if $|PA_{ERR}|> 90^\circ$ and $Q < 0.7$ as this combination indicates a poorly constrained position angle in a non-circularized object. $|PA_{ERR}|= 90^\circ$ is set for these objects upon flagging them. This flag indicates objects that have a reliable position angle measurement in at least 10 filters (1) and (0) otherwise; 153 objects fail this requirement and are assigned {\texttt{USE\_PA}} = 0. 

For each galaxy, we provide structural parameters measured in each wide- and medium-band JWST/NIRCam filter, including the $X$- and $Y$-centroid positions in pixels, AB magnitude, effective radius in arcseconds and kiloparsecs, circularized effective radius in kiloparsecs, Sérsic index $n$, axis ratio $Q$, and position angle $PA$. Physical sizes in kiloparsecs are computed from the arcsecond measurements using the maximum-likelihood photometric redshift $z_\mathrm{ml}$ from {\texttt{EAzY}}; this redshift is also used to derive rest-frame optical ($0.5\,\mu$m) and rest-frame NIR ($1.5\,\mu$m) sizes by selecting the available filter whose effective wavelength most closely corresponds to the target rest-frame wavelength at the redshift of each galaxy. We additionally provide the median size across all filters; we adopt the robust measure of biweight location to represent the median of the distribution in order to minimize the effects of outliers. All quoted sizes are best-fit models from \galfit\ which are PSF-deconvolved, and uncertainties are provided for each parameter. A full description of the catalogue columns is given in Appendix~\ref{app:cat-params}.

A companion catalogue with morphological parameters of the galaxies in the CANUCS cluster fields (CLU) from the combined CANUCS and JUMPS programs (JWST Ultimate Medium-band Photometric Survey; ID: 5890; PI: Withers, Muzzin) imaging will be made available by J. Judež et al. (in prep).

\subsection{Sample Selection for Multi-wavelength Size Analysis} \label{sec:data selection}

We study the multi-wavelength sizes of galaxies over the redshift range $0.6 < z_{\mathrm{phot}} \leq 4$, with the goal of investigating the rest-frame wavelength dependence of the galaxy size--mass relation out $z = 4$. We choose a minimum redshift constraint of $z = 0.6$ in order to avoid cluster infall from the MACS~J1149 and MACS~J1423 fields which are located at redshifts of $z \sim 0.54$. We begin with the parent sample of 41,202 galaxies described in Section~\ref{sec:data products} and apply a series of cuts to construct a robust multi-wavelength sample. 

First, we retain only galaxies for which the best-fit Sérsic index does not reach the \galfit\ bounds in any given filter, requiring $0.1 < n < 10$ as described in Section~\ref{subsec:galfit}. For this analysis, we additionally require a photometric signal-to-noise ratio of $S/N > 20$ (calculated using Kron fluxes; \citealt{Sarrouh2025}) in at least five available filters to ensure that each galaxy has sufficient morphological information for a reliable multi-wavelength analysis (e.g. \citealt{vanderWel2012_snr, miller2024jwstuncoversopticalsize, miller2026bandiirelationshipoptical}). Together, these cuts reduce the parent sample to 6,719 galaxies.

Galaxy structural and star-forming scaling relations vary dramatically with star-formation activity (e.g. \citealt{barro-scaling-relations, bezanson-scaling-relations, Ward_2024}), and it is therefore necessary to separate star-forming and quiescent galaxies in order to independently derive relations reflective of their distinct evolutionary pathways.  We opt for classification via rest-frame UVJ colour selection, with rest-frame colours determined using \eazy\ \citep{Brammer2008-eazy}. Quiescent galaxies are identified following the prescription of \citet{Antwi-Danso_2023}, with selection criteria calibrated for the CANUCS fields for $0.5 \leq z \leq 6$,

\begin{figure}
    \centering
    \includegraphics[width=\linewidth]{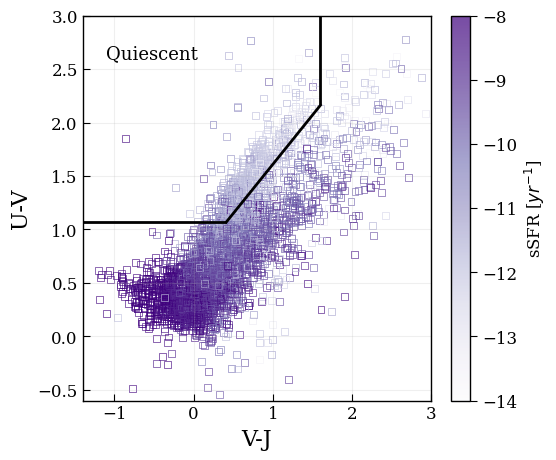}
    \caption{Rest-frame UVJ colour--colour diagram for the sample following the quality cuts described in Section~\ref{sec:data selection}, with specific star formation rate from \db\ fits (Section \ref{subsec:sed fitting}). Quiescent galaxies are identified using the criteria of \citealt{Antwi-Danso_2023} with selection criteria calibrated for the CANUCS fields for $0.5 \leq z \leq 6$ (Equation~\ref{eq:uvj_selection}).}
    \label{fig:uvj}
\end{figure} 

\begin{equation}
    \begin{array}{rcl}
    U - V &>& 1.07 \\
    V - J &<& 1.60 \\
    U - V &>& 0.92\,(V - J) + 0.69
    \end{array}
    \label{eq:uvj_selection}
\end{equation}
and all remaining galaxies are classified as star-forming. Figure~\ref{fig:uvj} shows the resulting UVJ distribution of the sample, with quiescent and star-forming galaxies indicated. This selection yields 563 quiescent and 6,138 star-forming galaxies with a quiescent fraction of $~\sim 8.5\%$, slightly lower than predicted compared to expectations of $\sim 10-20\%$ quenched up to Cosmic Noon, $z = 3.5$ \citep{Muzzin_2013}. The distributions of photometric redshift, stellar mass, and specific star formation rate for both populations are shown in Figure~\ref{fig:UVJ-params} (Appendix~\ref{app:num sources}). We note the limitations of uncorrected rest-frame colour estimation, including underestimated scatter and uncertainties (see \citealt{Noirot_2022}).

Since quiescent galaxies are systematically smaller and fainter than star-forming galaxies at fixed stellar mass \citep{bell-quiescent-vs-sf, deGraff-qui-vs-sf}, they have a higher mass-completeness limit and require separate treatment; therefore, we constrain our final sample to strictly star-forming objects. Mass-completeness limits for the CANUCS NCF fields are computed following the method of \citet{rosi_completeness}, which is briefly summarized in Appendix~\ref{app:mass-completeness}. We define redshift-evolving completeness limits as the stellar mass above which the sample is 50\% complete for star-forming galaxies in each redshift bin,
\begin{equation}
\log(M_\mathrm{complete}/M_\odot) =
    \begin{cases}
        8.0, & 0.6 < z \leq 1.0 \\
        8.5, & 1.0 < z \leq 1.5 \\
        8.5, & 1.5 < z \leq 2.0 \\
        9.0, & 2.0 < z \leq 4.0
    \end{cases}
    \label{eq:completeness}
\end{equation}
We adopt a relatively wide upper redshift bin of $2 < z \leq 4$ to maintain a sufficiently large sample while ensuring it spans roughly the same cosmic time interval as the lowest redshift bin, $0.6 < z \leq 1.0$. To ensure the robustness of results against incompleteness, a uniform lower mass limit of $\log(M_*/M_\odot) = 8.5$ is adopted across all redshift bins for this analysis, which corresponds to the most conservative completeness limit below $z = 2$. The sample spans up to $\log(M_*/M_\odot) \sim 11.5$, which we adopt as a representative upper limit. This yields a final sample of 4,140 star-forming galaxies across $0.6 < z \leq 4$. The distribution of galaxies across redshift and stellar mass bins is given in Table~\ref{tab:num-sources} (Appendix~\ref{app:num sources}).

\begin{figure*}
    \centering
    \includegraphics[width=0.99\textwidth]{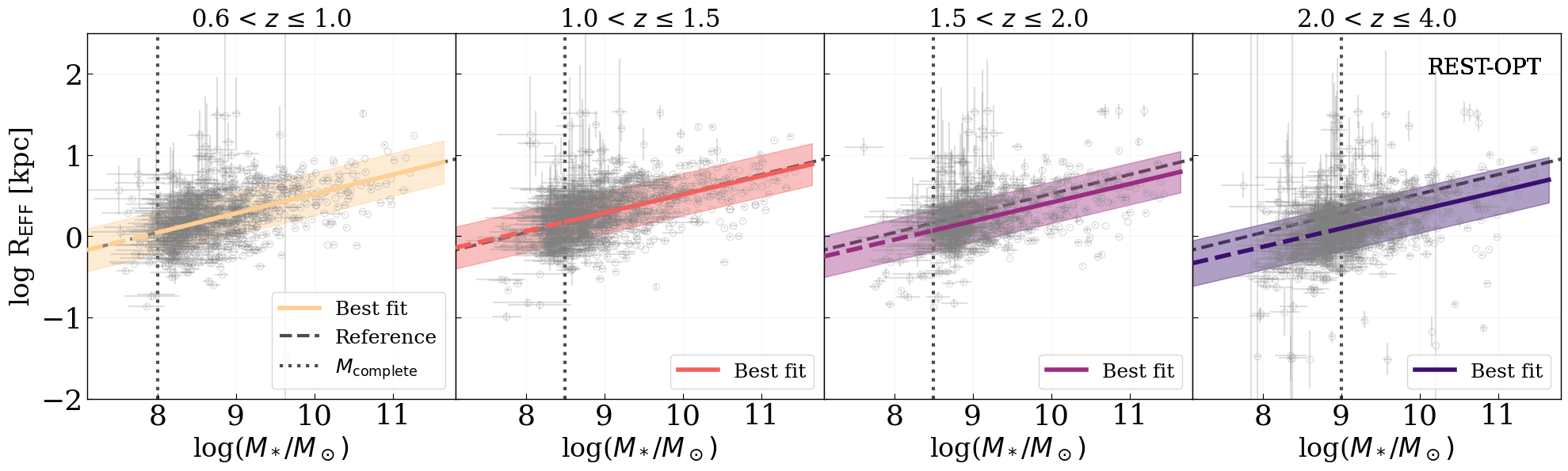}\\
    \includegraphics[width=0.99\textwidth]{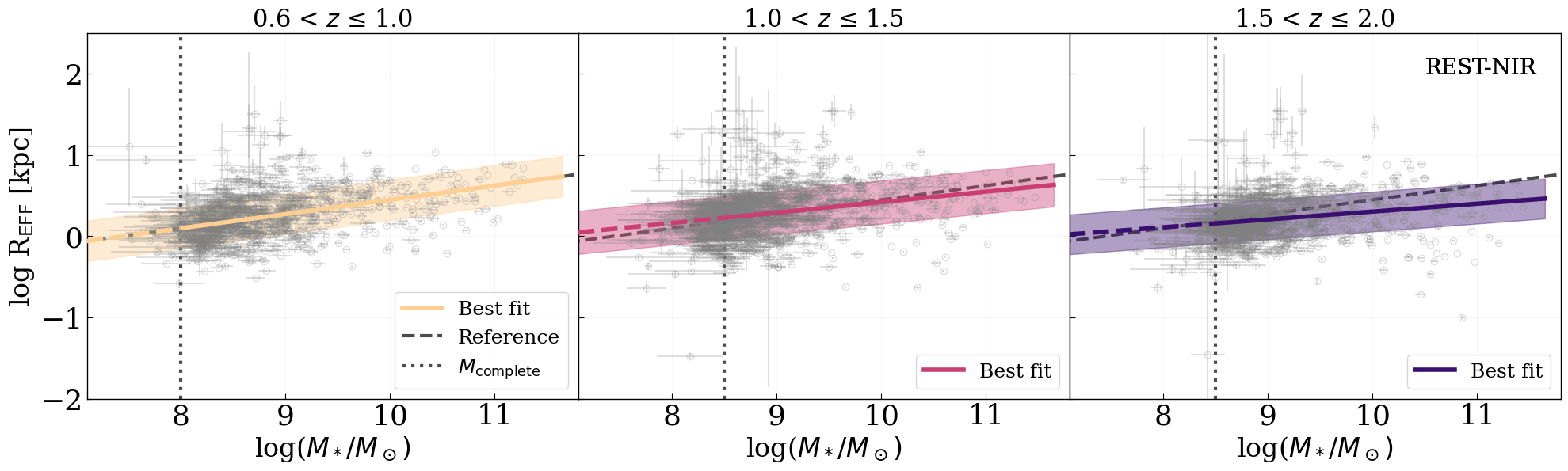}
    \caption{The size--mass relation of star-forming galaxies at the rest-frame 0.5$\mu m$ (top) and 1.5 $\mu$m (bottom) between $0.6 < z \leq 4$ and $8.5 < \log(M_*/M_\odot) \leq 11.5$. Data points are shown in grey with best-fits lines overplotted (with the dashed portion representing extrapolation below the mass completeness limit) and the $1\sigma$ intrinsic scatter is shaded. The black dashed line shows the slope of the lowest redshift bin, $0.6 < z \leq 1$, across redshift bins for reference. The vertical dotted line is the mass completeness limit for each bin (Equation \ref{eq:completeness}).}
    \label{fig:rf 0.5/1.5 best-fits}
\end{figure*}

\subsection{Size Determination with LinMix}\label{linmix}
We fit the size--mass relation between $0.6 < z \leq 4$ using the Bayesian linear regression algorithm \texttt{LinMix} \citep{kellylinmix}, following the approach of \citealt{Allen2024}. The relation is parametrized as,
\begin{align}\label{eq:size-mass}
\log\!\left(\frac{R_{\mathrm{eff}}}{\mathrm{kpc}}\right) = \alpha\,\log_{10}(M_*/M_\odot) + \log_{10}(A),
\end{align}
where $\alpha$ is the slope and $\log_{10}(A)$ in kpc is the intercept. \texttt{LinMix} is run with a default of $K = 3$ Gaussian components in the mixture model and a minimum of 100 MCMC iterations, and outputs posterior chains for the slope $\alpha$, intercept $\log_{10}(A)$, and intrinsic scatter $\sigma_{\log R_\mathrm{eff}}$. The best-fit parameters and their uncertainties are taken as the median and standard deviation of the respective chains. 

\section{Results}\label{results}
In this section, we present a multi-wavelength analysis of the star-forming galaxy size--mass relation over $0.6 < z \leq 4$ and $8.5 < \log(M_*/M_\odot) \leq 11.5$, enabled by the unprecedented sensitivity and wavelength coverage of JWST from the rest-frame UV to the rest-frame NIR. We present three main results: first, in Section~\ref{sec:rf_sizemass}, we present the rest-frame optical ($0.5\,\mu$m) and rest-frame NIR ($1.5\,\mu$m) size--mass relations, fit using Bayesian linear regression, and compare our results to existing literature across a broad redshift baseline. In Section~\ref{sec:size_wav}, we isolate the effects of stellar mass and redshift on the size--wavelength relation by holding each variable fixed in turn, and examine how the slope of the size--wavelength relation evolves across mass and redshift bins. Finally, in Section~\ref{sec:size_mass_wav}, we introduce stellar mass, redshift, and rest-frame wavelength simultaneously as free parameters in a weighted least-squares regression framework, and derive a generalized parametrization of galaxy size as a continuous function of all three variables between $0.25\mu m < \lambda_\mathrm{RF} < 2.0\mu m$.

\begin{table*}[ht!]
    \centering
    \begin{tabular}{cc|ccc|ccc}
        \hline\hline
        Redshift bin & $z_{\rm med}$ 
        & \multicolumn{3}{c|}{$\lambda_{\rm RF} = 0.5\,\mu$m} 
        & \multicolumn{3}{c}{$\lambda_{\rm RF} = 1.5\,\mu$m} \\
        & 
        & $\alpha$ & $\log_{10}(A)$ & $\sigma$ 
        & $\alpha$ & $\log_{10}(A)$ & $\sigma$ \\
        \hline
        $(0.6, 1.0]$ & 0.81 
        & $0.238 \pm 0.015$ & $-1.858 \pm 0.129$ & $0.264 \pm 0.006$ 
        & $0.176 \pm 0.012$ & $-1.313 \pm 0.107$ & $0.250 \pm 0.006$ \\
        
        $(1.0, 1.5]$ & 1.21
        & $0.225 \pm 0.015$ & $-1.733 \pm 0.138$ & $0.259 \pm 0.007$ 
        & $0.132 \pm 0.013$ & $-0.890 \pm 0.118$ & $0.266 \pm 0.006$ \\
        
        $(1.5, 2.0]$ & 1.73 
        & $0.228 \pm 0.015$ & $-1.863 \pm 0.144$ & $0.255 \pm 0.008$ 
        & $0.095 \pm 0.013$ & $-0.653 \pm 0.118$ & $0.244 \pm 0.006$ \\
        
        $(2.0, 4.0]$ & 2.54 
        & $0.225 \pm 0.022$ & $-1.929 \pm 0.210$ & $0.282 \pm 0.008$ 
        & -- & -- & -- \\
        \hline
    \end{tabular}
    \caption{Best-fit parameters of the size--mass relation from \texttt{LinMix} for each redshift evaluated at the rest-frame $0.5\,\mu$m and $1.5\,\mu$m reporting slope $\alpha$, normalization $\log_{10}(A)$, and intrinsic scatter $\sigma$. The median redshift of each bin is also listed.}
    \label{tab:linmix-params}
\end{table*}

\subsection{Rest-Frame Optical and NIR Size--Mass Relation}\label{sec:rf_sizemass}
We present the size--mass relation measured at the rest-frame optical ($0.5\,\mu$m) and rest-frame NIR ($1.5\,\mu$m) for star-forming galaxies over $0.6 < z \leq 4$, following the method of Section \ref{linmix}. For each galaxy, we select the JWST/NIRCam filter whose effective wavelength most closely corresponds to the target rest-frame wavelength. The rest-frame $1.5\,\mu$m corresponds to an observed wavelength of $\gtrsim 4.5\,\mu$m at $z > 2$, which falls beyond the wavelength coverage of the longest available JWST/NIRCam filter in all five fields, F444W ($\lambda_\mathrm{eff} = 4.4\,\mu$m). The rest-frame NIR size--mass relation is therefore presented only for $0.6 < z \leq 2$. 

\begin{figure}[htp!]
    \centering
    \includegraphics[width=\linewidth]{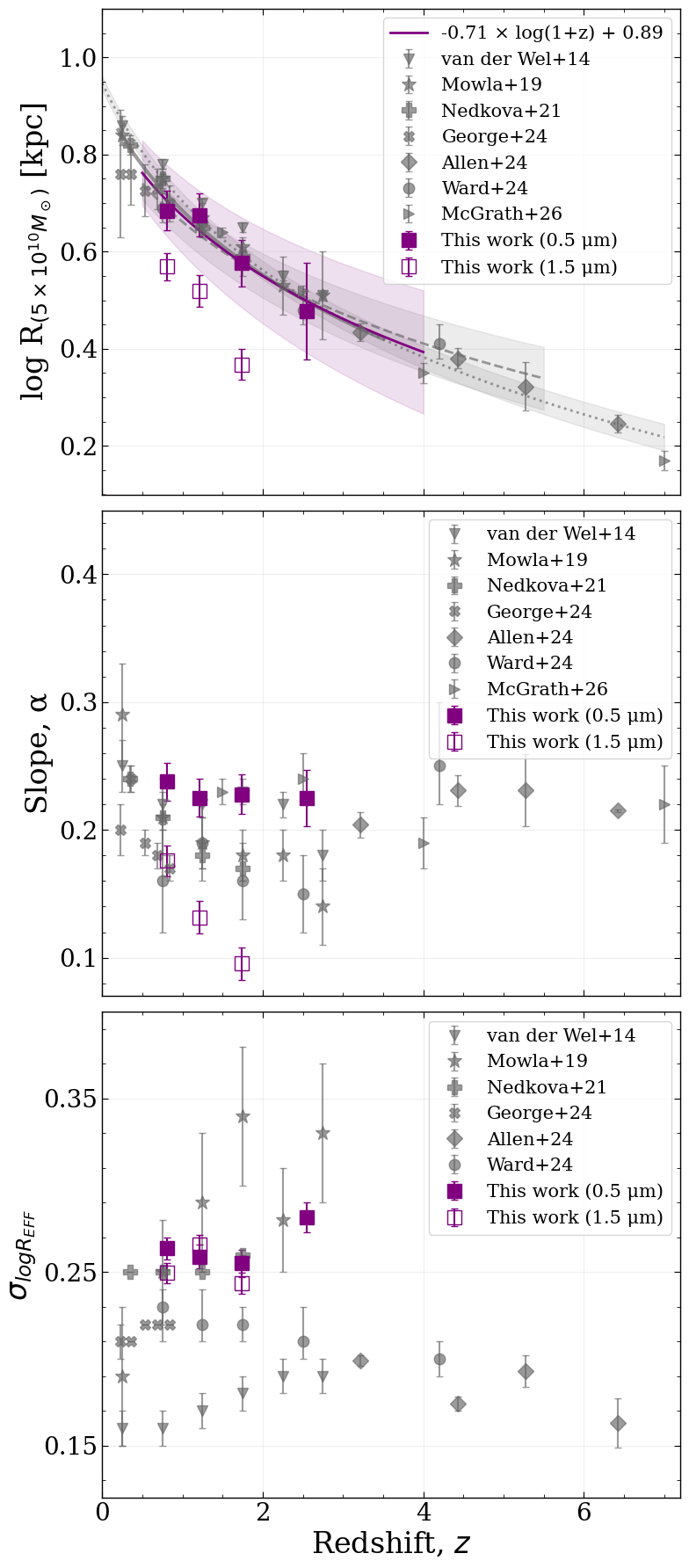}
    \caption{Evolution of the rest-frame optical size--mass relation parameters for star-forming galaxies at $0.6 < z \leq 4$. Top (a): rest-optical effective radius at a fixed stellar mass of $M_* = 5 \times 10^{10}\,M_\odot$ as a function of redshift, compared to literature values from \citet{vanderWel_2014}, \citet{Mowla_2019}, \citet{nedkova2021}, \citet{Ward_2024}, \citet{Allen2024}, \citet{George2024}, and \citet{McGrath_2026}. Middle (b): evolution of the size--mass slope $\alpha$. Bottom (c): evolution of the intrinsic scatter $\sigma_{\log R_\mathrm{eff}}$. Rest-NIR values from this work are shown for comparison in all panels.}
    \label{fig:combined}
\end{figure}

The resulting rest-frame optical and NIR size--mass relations, along with the best-fit lines and $1\sigma$ intrinsic scatter (shaded regions), are shown in Figure~\ref{fig:rf 0.5/1.5 best-fits} for all redshift bins. The slope of the lowest redshift bin, $0.6 < z \leq 1$, is shown as a reference to clearly demonstrate the evolution of the size--mass relation as a function of redshift. Corresponding fit parameters are listed in Table \ref{tab:linmix-params}. 
At the rest-frame optical, the slope of the size--mass relation is in effect preserved across redshift; the average size of galaxies decreases at higher redshifts, as expected, but size as a function of mass follows a systematic scaling up to $z = 4$ with relatively constant values of $\alpha$ (see Table \ref{tab:linmix-params}). The reported slopes of the rest-optical size--mass relation for star-forming galaxies, $\alpha \sim 0.2$, are consistent with literature values \citep{vanderWel_2014, Mowla_2019, Kawinwanichakij_2021, nedkova2021, McGrath_2026}. At the rest-frame NIR, the slope of the size--mass relation flattens moderately as a function of redshift with a $\sim 50\%$ decrease in $\alpha$ between $0.6 < z \leq 1$ and $1.5 < z \leq 2$, with more pronounced deviation on the high-mass end.
The size evolution at a fixed stellar mass of $M_* = 5 \times 10^{10}\,M_\odot$ is fit with the relation,
\begin{align}
\log R_{\mathrm{eff}} = \beta_z \times \log_{10}(1+z) + B_z,
\end{align}
using a first-degree polynomial weighted by the uncertainty in size. For the rest-frame optical, we obtain,
\begin{align}
\log R_{\mathrm{eff, opt}} = (-0.71 \pm 0.16)\times\log_{10}(1+z) + (0.89 \pm 0.06), 
\end{align}
for $0.6 < z \leq 4$. Figure~\ref{fig:combined}a shows the rest-optical size evolution in comparison to literature values from \citet{Allen2024}, \citet{Ward_2024}, \citet{Mowla_2019}, and \citet{vanderWel_2014}, with which our results are in agreement within uncertainties. The evolution of the slope $\alpha$ and intrinsic scatter $\sigma_{\log R_\mathrm{eff}}$ with redshift are shown in Figures~\ref{fig:combined}b and~\ref{fig:combined}c, respectively, alongside rest-NIR values from this work for comparison. Our rest-optical slope values (shown in purple) are highly consistent with previous literature at $z < 4$ (e.g. \citealt{vanderWel_2014, Mowla_2019, Kawinwanichakij_2021, nedkova2021, McGrath_2026}), with the average $\alpha_{0.5\mu m} = 0.230 \pm 0.008$. 


\begin{figure}[htp!]
\centering
\includegraphics[width=\linewidth]{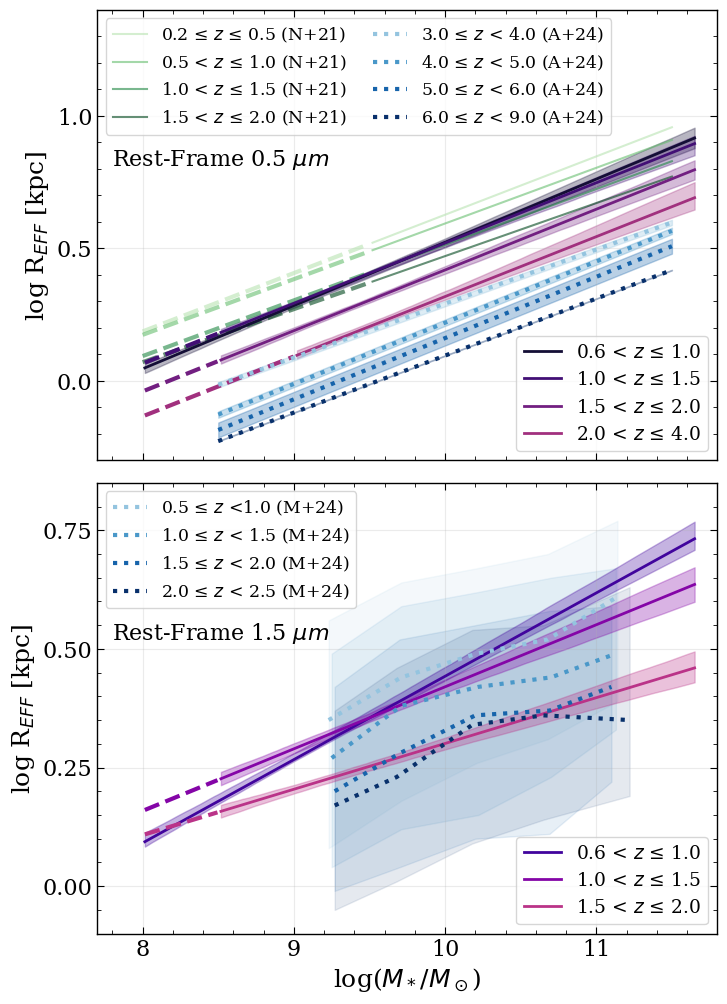}
\caption{Best-fit size--mass relations for star-forming galaxies in each redshift bin at rest-frame $0.5\,\mu$m (top) compared to \citealt{nedkova2021} (N+21) and \citealt{Allen2024} (A+24), and at the rest-frame $1.5\,\mu$m (bottom) compared to \citealt{Martorano_sizemass} (M+24). Dashed lines represent extrapolations and shaded regions indicate the $1\sigma$ intrinsic scatter.}
\label{fig:linear-size-mass}
\end{figure}

Figure~\ref{fig:linear-size-mass} shows the best-fit size--mass relations at rest-frame $0.5\,\mu$m and $1.5\,\mu$m in each redshift bin, compared directly to both HST and JWST results. At the rest-frame $0.5\,\mu$m, a self-consistent result is observed, with a uniform decrease in galaxy size with increasing redshift and a constant slope across redshift bins and mass ranges probed; the derived slopes are consistent compared to literature with $\Delta~\alpha \lesssim 0.04$ from the relations reported (\citealt{nedkova2021} and \citealt{Allen2024}).

At the rest-frame $1.5\,\mu$m, our results are in agreement within $1\sigma$ of \citet{Martorano_sizemass} at the mass ranged probed, $\log(M_*/M_\odot) \gtrsim 9.2$; a shallower slope is reported by \citet{Martorano_sizemass} at comparable redshifts $\sim15\% $ decrease in the average slope across redshift bins. This likely reflects limitations in low-mass coverage; a detailed investigation of the size--mass slope at $\log(M_*/M_\odot) \lesssim 9.5$ is deferred to Section~\ref{discussion}. 

\begin{figure*}
    \gridline{
        \fig{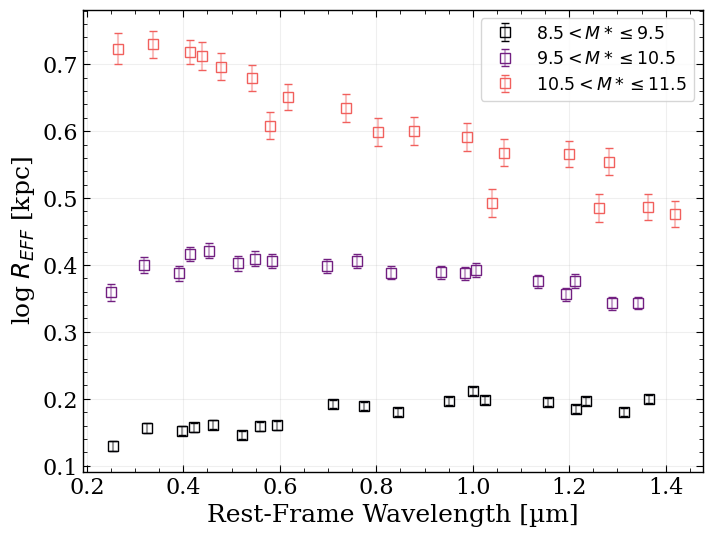}{0.48\textwidth}{(a)}
        \fig{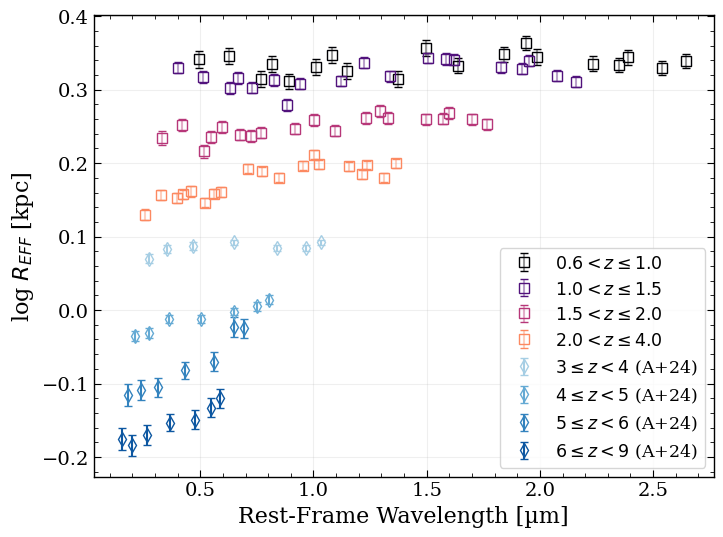}{0.49\textwidth}{(b)}
    }
    \gridline{
        \fig{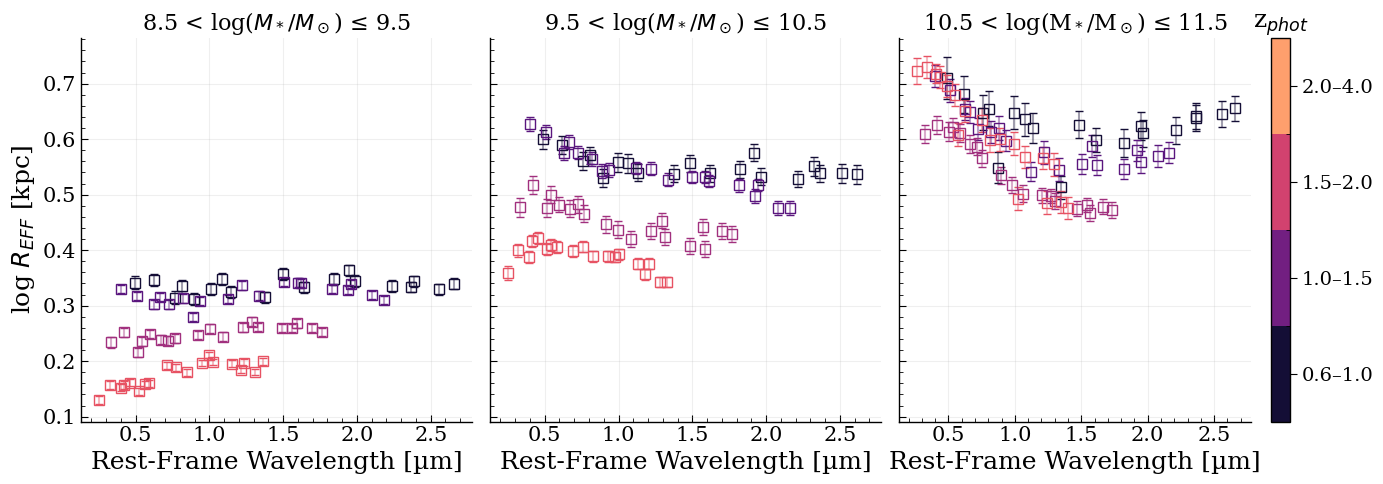}{\textwidth}{(c)}
    }
    \caption{Top (a): the size--wavelength relation as a function of stellar mass at fixed redshift ($2 < z \leq 4$), colour-coded by mass bin. Bottom (b): the size--wavelength relation as a function of redshift at fixed stellar mass ($8.5 \leq \log(M_*/M_\odot) < 9.5$), colour-coded by redshift bin and compared to \citet{Allen2024} at $z \geq 3$. Panel (c): weighted average size--wavelength relations for star-forming galaxies at $0.6 < z \leq 4$ and $8.5 < \log(M_*/M_\odot) \leq 10.5$ and $10.5 < \log(M_*/M_\odot) \leq 11.5$.}
    \label{fig:avg_fits-CHEBY}
\end{figure*}

\subsection{Size--Wavelength Relation by Redshift and Mass}\label{sec:size_wav}
We present a systematic multi-band analysis of the size--wavelength relation spanning the full range of available JWST medium- and wide-band filters extending down to $\sim 10^{8.5} M_\odot$. By the following methods, we isolate each of these variables in turn before combining them in a unified regression framework in Section~\ref{sec:size_mass_wav}.

To test the dependence of the size--wavelength relation on stellar mass, we fix the redshift to $2 < z \leq 4$, the most populated redshift bin in our sample, and compute the \galfit\ error-weighted average size as a function of rest-frame wavelength in three stellar mass bins: $\log(M_*/M_\odot) = 8.5$--$9.5$, $9.5$--$10.5$, and $10.5$--$11.5$. This weighted-averaging approach follows \citet{R-Allen-size-mass} and \citet{Allen2024}, and the resulting size--wavelength relations are shown in Figure~\ref{fig:avg_fits-CHEBY}a. A clear trend is observed in which more massive galaxies exhibit a steeper decline in size with increasing wavelength, while the lowest-mass bin ($8.5 < \log(M_*/M_\odot) \leq 9.5$) shows a markedly flatter size--wavelength slope. The linear best-fit slopes of the size--wavelength relation for the three mass bins are $\beta \approx$  0.05, -0.04, and -0.22 for $\log(M_*/M_\odot) = 8.5$--$9.5$, $9.5$--$10.5$, and $10.5$--$11.5$, respectively, confirming that the slope of the size--wavelength relation decreases with increasing stellar mass.

To isolate the redshift dependence, we fix the stellar mass to $8.5 \leq \log(M_*/M_\odot) < 9.5$, consistent with the mass normalization adopted by \citet{Allen2024}, and compute the weighted average size--wavelength relation in our respective redshift bins. Rest-frame wavelengths are computed using the median redshift of each bin ($z_\mathrm{med} = 0.81$, $1.21$, $1.73$, and $2.54$, respectively), and the results are shown in Figure~\ref{fig:avg_fits-CHEBY}b alongside the high-redshift measurements of \citet{Allen2024} for comparison. The size--wavelength slope is notably flat at $z \lesssim 4$ and shows little variation across the redshift bins probed in this work, suggesting that redshift plays a secondary role compared to stellar mass in driving the wavelength dependence of galaxy size. At $z > 5$, \citet{Allen2024} report a tentative increase in the size--wavelength slope with redshift; two potential interpretations include: stronger colour gradients at higher redshift, arising from the more centrally concentrated star formation in galaxies at earlier evolutionary stages, or an increased contribution from dust attenuation in the rest-NIR, which preferentially affects the high-$z$ size--wavelength relations of \citet{Allen2024} that probe closer to the rest-optical ($\sim 0.5\,\mu$m). The role of dust and colour gradients in driving these trends is examined in detail in Section~\ref{discussion}.

\subsection{Size--Mass--Wavelength Relation}\label{sec:size_mass_wav}
Having established the independent effects of stellar mass and redshift on the size--wavelength relation in Section~\ref{sec:size_wav}, we now introduce both variables simultaneously as free parameters. Figure~\ref{fig:avg_fits-CHEBY}c shows the \galfit\ error-weighted average size--wavelength relations computed for star-forming galaxies across $0.6 < z \leq 4$ and $8.5 < \log(M_*/M_\odot) \leq 11.5$. A clear and systematic trend in the size--wavelength slope as a function of stellar mass is apparent: higher-mass galaxies exhibit a steeper decline in size with increasing wavelength, while lower-mass galaxies show a comparatively flat size--wavelength relation independent of redshift up to $z=4$. We note the limited sample size of the highest mass bin ($\lesssim 75$ galaxies per redshift bin in $10.5 < \log(M_*/M_\odot) \leq 11.5$), which likely drives the higher uncertainty and scatter in the trend. 

\begin{figure*}
\centering
\includegraphics[width=\linewidth]{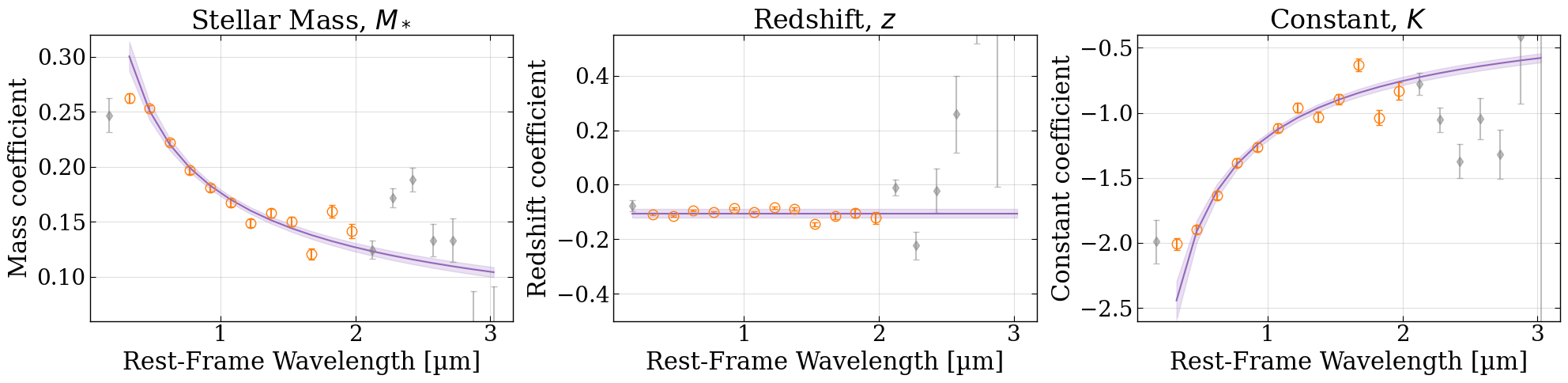}
\caption{Best-fit coefficients of the weighted least-squares regression as a function of rest-frame wavelength bin median: the mass coefficient $m(\lambda)$ (left), the redshift coefficient $z(\lambda)$ (center), and the constant term $K(\lambda)$ (right), with $1\sigma$ uncertainties. Purple lines show the best-fit power-law (mass and constant terms) and linear (redshift term) functional forms over $0.25\mu m < \lambda_\mathrm{RF} < 2.0\mu m$ with a $1\sigma$ confidence interval (shaded). Grey points fall outside this range and are excluded from the functional fits due to an insufficient sample size at wavelength extremes.}
\label{fig:coeffs_fits}
\end{figure*}

The observed trend is quantified by a Spearman rank correlation coefficient of $\rho_{M_*} = -0.47$ between stellar mass and the size--wavelength slope, confirming a moderately inverse relationship. In contrast, the corresponding correlation with redshift is negligible ($\rho_z = -0.01$), consistent with the finding in Section~\ref{sec:size_wav} that redshift plays a secondary role to mass in driving wavelength-dependent size variations. These results are also consistent with previous high-mass studies \citep{vanderWel_2014, Kawinwanichakij_2021, jia_2024_size_wav, George2024, miller2026bandiirelationshipoptical}; in particular, \citet{jia_2024_size_wav} and \citet{George2024} present that star-forming galaxies at $z < 3.5$ and $z < 1$ respectively are systematically smaller at longer wavelengths, with the effect most pronounced at higher stellar masses.

Notably, the lowest mass bin ($8.5 < \log(M_*/M_\odot) \leq 9.5$) displays a markedly flatter size--wavelength slope relative to higher-mass bins, with galaxy sizes showing little to no wavelength dependence across the full rest-frame optical to NIR range. This behaviour suggests that $\log(M_*/M_\odot) \sim 9$--$9.5$ may represent a transitional mass scale at which the dominant structural and star-formation processes governing galaxy morphology change character. We return to this point in the context of the full multi-wavelength size--mass relation in Section~\ref{discussion}.

To derive a generalized parametrization of galaxy size as a function of rest-frame wavelength, stellar mass, and redshift simultaneously, we perform a weighted least-squares (WLS) regression on the full size--wavelength dataset. Galaxies are binned by rest-frame wavelength in intervals of $0.15\,\mu$m over $\lambda_\mathrm{RF} = 0.1$--$3\,\mu$m (see Appendix~\ref{wav-bins} for bin counts per wavelength range), and each galaxy is weighted by its \galfit\ uncertainty as defined in Section~\ref{subsec:galfit}. The regression is performed using \texttt{statsmodels} \citep{seabold2010} with stellar mass, photometric redshift, and a constant intercept term as free parameters in each wavelength bin. 
Figure~\ref{fig:coeffs_fits} shows the best-fit coefficients for the mass term $m(\lambda)$, the redshift term $z(\lambda)$, and the constant term $K(\lambda)$ as a function of rest-frame wavelength. To reduce the influence of poorly populated bins at the extremes of the wavelength range, we conservatively restrict the functional parametrization to $0.25\mu m < \lambda_\mathrm{RF} < 2.0\mu m$, where the sample is well-characterized (see Appendix~\ref{wav-bins}). The mass and constant coefficients exhibit a smooth power-law dependence on wavelength, and are fit with a functional form $a \times \lambda^b$. The redshift coefficient shows no significant wavelength evolution over this range and is therefore modelled as a constant; the derived best-fit functional forms are,
\begin{align}
m(\lambda) &= 0.18 \pm 0.003 \times \lambda^{-0.47 \pm 0.04},\\
z(\lambda) &= -0.11 \pm 0.02,\\
K(\lambda) &= -1.2 \pm 0.03 \times \lambda^{-0.65 \pm 0.05}.
\end{align}

Combining these, the generalized size--mass--wavelength relation for star-forming galaxies over $0.25\mu m < \lambda_\mathrm{RF} < 2.0\mu m$ is,
\begin{align}
\log\left(\frac{R_\mathrm{eff}}{\mathrm{kpc}}\right) = m(\lambda)\,\log(M_*/M_\odot) + z(\lambda)\,z_\mathrm{phot} + K(\lambda),
\label{eq:FINAL}
\end{align}
given by,
\begin{align*}
\log\left(\frac{R_\mathrm{eff}}{\mathrm{kpc}}\right) =~ &(0.18 \pm 0.003 \times \lambda^{-0.47 \pm 0.04})\,\log\left(\frac{M_*}{M_\odot}\right) &+\\[-0pt]
&(-0.11 \pm 0.02)\,z_\mathrm{phot} &+\\
&(-1.2 \pm 0.03 \times \lambda^{-0.65 \pm 0.05})
\end{align*}

This relation reproduces observed galaxy sizes with a residual scatter of $\sim 0.05$\,dex (Appendix~\ref{app:residuals}).

\subsection{Multi-Wavelength Size--Mass Relation}\label{sec:multi-wav-sizemass}
To construct a multi-wavelength size--mass relation, we select the deepest available filters with a $5\sigma$ point-source depth of $f < 4.5$\,nJy , comprising of F090W, F115W, F150W, F182M, F200W, F277W, F335M, F356W, F360M, F410M, and F444W, to ensure the highest $S/N$ morphological measurements across the available wavelength range. The redshift-evolving stellar mass completeness limits from Equation~\ref{eq:completeness} are applied. 

\begin{figure*}[htp!]
\centering
\includegraphics[width=\linewidth]{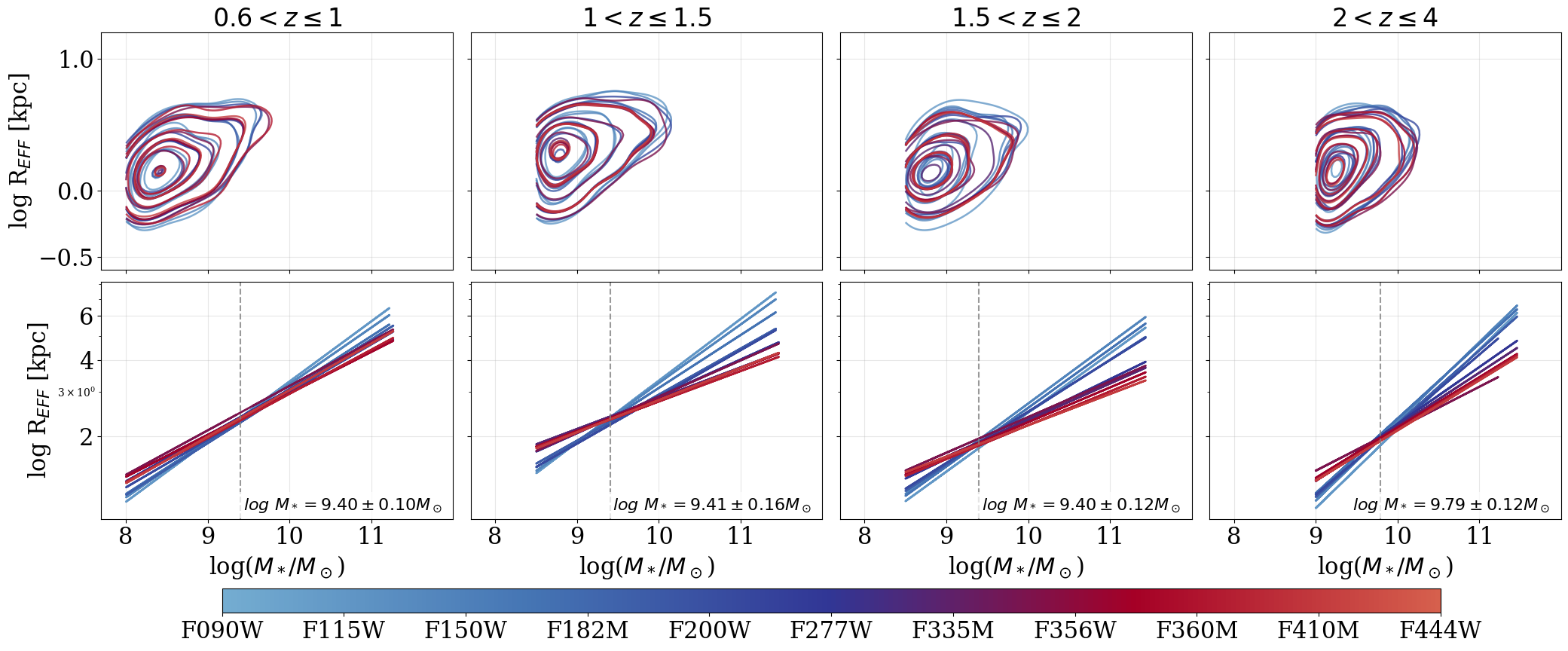}
\caption{The multi-wavelength size--mass relation for star-forming galaxies at $0.6 < z \leq 4$, shown in each of the 11 selected JWST/NIRCam filters colour-coded by wavelength. Upper panels show the density contours of individual best-fit size--mass measurements; lower panels show the \texttt{LinMix} linear best-fits in each redshift bin. The dashed vertical line indicates the characteristic mass per redshift bin at which the size--mass relations measured across different filters converge.}
\label{fig:size-mass FINAL}
\end{figure*}

Figure~\ref{fig:size-mass FINAL} presents the multi-wavelength size--mass relation for star-forming galaxies at $0.6 < z \leq 4$, shown separately for each filter (top panels) and as linear best-fits in each redshift bin (bottom panels). The fits are obtained using the \texttt{LinMix} Bayesian regression procedure described in Section~\ref{sec:rf_sizemass}, applied independently to each filter; slope and intercept values for each redshift bin and filter are presented in Table \ref{tab:size_mass_slopes_intercepts}. A clear and systematic gradient in the size--mass slope is observed as a function of rest-frame wavelength, with the NIR size--mass relation exhibiting a consistently shallower slope than the rest-frame optical across all redshift bins. This trend is consistent with previous multi-wavelength studies of the size--mass relation (e.g. \citealt{Suess_2022sizemass, Martorano_sizemass, vdW_2024_sizemass, miller2026bandiirelationshipoptical}).

\begin{table*}[ht!]
    \centering
    \begin{tabular}{lcccc|cccc}
    \hline\hline
    Filter & \multicolumn{4}{c}{Slope} & \multicolumn{4}{c}{Intercept} \\
    \hline
    & $(0.6-1.0]$ & $(1.0, 1.5]$ & $(1.5, 2.0]$ & $(2.0, 4.0]$ 
    & $(0.6-1.0]$ & $(1.0, 1.5]$ & $(1.5, 2.0]$ & $(2.0, 4.0]$  \\
    \hline
    F090W & 0.239 & 0.245 & 0.235 & 0.315 & -1.873 & -1.932 & -1.954 & -2.822 \\
    F115W & 0.225 & 0.233 & 0.239 & 0.315 & -1.743 & -1.816 & -1.964 & -2.790 \\
    F150W & 0.209 & 0.210 & 0.233 & 0.297 & -1.603 & -1.611 & -1.918 & -2.598 \\
    F182M & 0.206 & 0.183 & 0.208 & 0.291 & -1.580 & -1.364 & -1.689 & -2.562 \\
    F200W & 0.189 & 0.186 & 0.204 & 0.278 & -1.419 & -1.399 & -1.646 & -2.438 \\
    F277W & 0.182 & 0.146 & 0.158 & 0.226 & -1.339 & -1.006 & -1.212 & -1.916 \\
    F335M & 0.167 & 0.122 & 0.145 & 0.213 & -1.194 & -0.773 & -1.086 & -1.787 \\
    F356W & 0.177 & 0.144 & 0.142 & 0.167 & -1.266 & -0.981 & -1.041 & -1.338 \\
    F360M & 0.165 & 0.124 & 0.138 & 0.200 & -1.184 & -0.798 & -1.025 & -1.668 \\
    F410M & 0.175 & 0.121 & 0.134 & 0.203 & -1.284 & -0.772 & -0.998 & -1.700 \\
    F444W & 0.185 & 0.128 & 0.124 & 0.199 & -1.366 & -0.838 & -0.899 & -1.665 \\
    \hline
    \multicolumn{1}{l}{\# Galaxies} & 903 & 1042 & 948 & 1027 &  --&  --&  --& -- \\
    \multicolumn{1}{l}{log($M_{complete}/M_\odot$)} & 8.0 & 8.5 & 8.5 & 9.0 &  --&  --&  --&  --\\
    \hline
    \end{tabular}
    \caption{Linear best-fits parameters for the star-forming size–mass relations by filter and redshift bin. The number of galaxies and the mass completeness limit per redshift bin are also given.}
    \label{tab:size_mass_slopes_intercepts}
\end{table*}

A notable feature of Figure~\ref{fig:size-mass FINAL} is the presence of a characteristic ``crossover'' mass, indicated by the dashed vertical line in each panel, at which the variance in size across filters is minimized and the size--mass relations measured at different wavelengths converge. We measure a mean crossover mass of $\log(\bar{M}_*/M_\odot) = 9.50 \pm 0.13$ averaged across all four redshift bins. A slight increase to $\log(M_*/M_\odot) \simeq 9.79 \pm 0.12$ is observed in the $2 < z \leq 4$ bin, which we attribute to the higher mass-completeness limit adopted at these redshifts ($\log M_\mathrm{complete}/M_\odot = 9.0$) rather than a genuine physical evolution. The uncertainty on each crossover mass is taken as the standard deviation in size across wavelength-dependent fits evaluated at this stellar mass.

\begin{figure}[ht!]
\centering
\includegraphics[width=\linewidth]{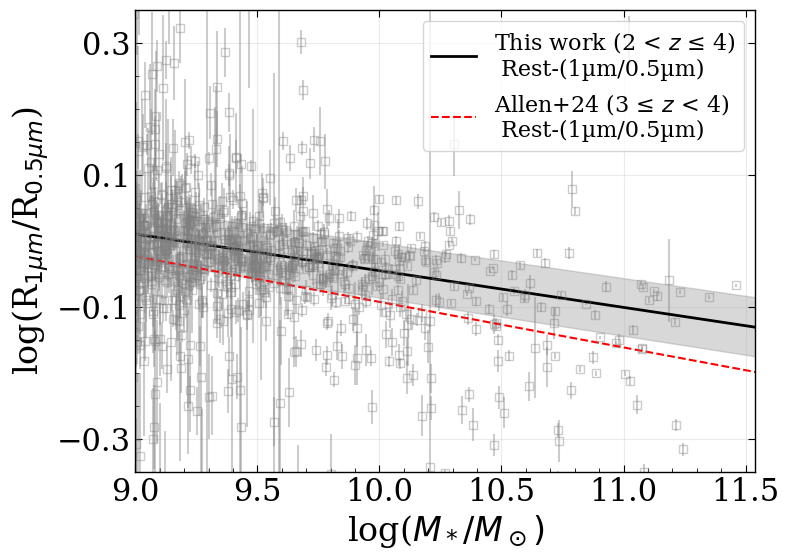}
\caption{Ratio of the rest-frame $1\,\mu$m to rest-frame $0.5\,\mu$m effective radius as a function of stellar mass, measured using F360M and F182M respectively. A ratio below unity indicates that galaxies are more extended at shorter wavelengths.}
\label{fig:size-mass ratio}
\end{figure}

To quantify the wavelength dependence of galaxy size above and below the crossover mass, we compute the ratio of the rest-frame $1\,\mu$m size to the rest-frame $0.5\,\mu$m size as a function of stellar mass, using the F360M and F182M filters as proxies for these rest-frame wavelengths, respectively. The resulting size ratio is shown in Figure~\ref{fig:size-mass ratio}. Above the crossover mass ($\log(M_*/M_\odot) \gtrsim 9.5$), galaxies are on average $10.48\%$ smaller at rest-frame $1\,\mu$m than at rest-frame $0.5\,\mu$m. This size difference increases to $26.74\%$ at $\log(M_*/M_\odot) > 10.5$, reflecting the growing influence of colour gradients at high stellar masses. Below the crossover mass ($\log(M_*/M_\odot) < 9.5$), there is a 2.05\% difference between rest-frame $1\mu m$ and rest-frame $0.5\mu m$ sizes, indicating a negligible effect from colour gradients on the low-mass end.

\section{Discussion}\label{discussion}
The multi-wavelength size--mass relation presented in Section~\ref{sec:multi-wav-sizemass} reveals a systematic and physically informative gradient in the size--mass slope as a function of rest-frame wavelength, with a characteristic crossover mass at $\log(M_*/M_\odot) \sim 9.5$ separating two distinct regimes of wavelength-dependent morphology. In this section, we interpret these trends in terms of the physical processes governing galaxy structure, including colour gradients, dust attenuation, and the spatial distribution of stellar populations. We discuss the implications for inside-out and outside-in modes of galaxy formation and compare our results to predictions from hydrodynamical simulations.

\subsection{The Origin of the Size--Wavelength Slope: Colour Gradients and Dust Attenuation}\label{sec:colour-gradients}
The systematic decrease in the size--mass slope with increasing rest-frame wavelength observed in Figure~\ref{fig:size-mass FINAL} is a manifestation of radial colour gradients within galaxies, in which the central and outskirt regions differ in their stellar population age, metallicity, and dust content \citep{dejong1996, Gadotti_2001, macarthur_2003}. In the rest-frame UV and optical, galaxy sizes are sensitive to regions of active star formation and to dust attenuation, both of which are typically more prominent at shorter wavelengths. In the rest-frame NIR, the light traces the older, more evolved stellar population and is significantly less affected by dust, providing a more direct proxy for the underlying stellar mass distribution \citep{Bell_ML_ratio, vanderWel_2014}.

Radial colour gradients in star-forming galaxies are well established observationally, with galaxies generally becoming redder towards their center (e.g. \citealt{Suess_2019, Allen2024, miller2024jwstuncoversopticalsize, Martorano2026}); they are naturally expected in the context of inside-out growth, where the central regions build up stellar mass earlier and host older, more metal-rich populations, while the outer disk continues to form stars and are thus bluer than the center (e.g. \citealt{Mo_1998, vanDokkum_2013}). Colour gradients directly produce wavelength-dependent size estimates; at UV/optical wavelengths, the extended star-forming disk contributes more flux and the measured size is larger, while at NIR wavelengths the compact, evolved central component dominates and the measured size is smaller. This is consistent with the trend observed above the crossover mass in our sample.

Dust attenuation provides a further and compounding contribution to the observed size--wavelength trend, particularly at high-masses ($\log(M_*/M_\odot) \gtrsim 10$) for star-forming galaxies \citep{Martis2016}. As shown in Appendix~\ref{app:dust}, dust attenuation $A_V$ increases sharply between $9 < \log(M_*/M_\odot) < 10$, coinciding with the crossover mass identified in Section~\ref{sec:multi-wav-sizemass}. In massive star-forming galaxies, dust is preferentially concentrated in the central regions, where star formation rates and gas surface densities are highest (e.g. \citealt{Wuyts_2011, Nelson_2016, Tacchella_2018, Matharu_2023}). This centrally concentrated dust attenuates UV and optical light from the compact central star-forming regions, causing the galaxy to appear more extended at short wavelengths than it truly is (e.g. \citealt{Zapata_2019,Sun_2021, bodansky2026jwstalmarevealbuildstellar}). NIR wavelengths, being far less sensitive to dust attenuation, penetrate through to the compact central stellar component and yield a smaller measured size. Intrinsic colour gradients driven by dust attenuation therefore provide a natural and self-consistent explanation for the steeper size--wavelength slope observed at high stellar masses.

Previous work with HST has suggested that dust attenuation alone may account for the observed differences in rest-UV and rest-optical sizes \citep{nedkova2024uvcandelsroleduststellar}. The extensive NIR coverage provided by JWST/CANUCS enables a more direct probe of the stellar mass distribution at wavelengths where dust effects are substantially reduced, allowing us to disentangle the contributions of dust and stellar population gradients more cleanly than was possible with HST alone. Our results are broadly consistent with a dust-driven explanation at high masses, but the behaviour at low masses, discussed in Section~\ref{sec:inside-out}, points to a complementary role for structural evolution in driving the observed wavelength dependence. 

\subsection{Crossover Mass as a Transition in Galaxy Structure}\label{sec:turnover}
The stability of the characteristic crossover mass with redshift is itself a physically interesting result, suggesting that this stellar mass reflects a fundamental transition in galaxy structure rather than a redshift-dependent observational effect. If the crossover were driven purely by mass assembly, one might expect it to shift to lower masses at lower redshifts as galaxies grow. The absence of significant redshift evolution in the crossover mass over the range $0.6 < z \leq 4$ instead suggests that the transition reflects an intrinsic mass scale in galaxy formation, potentially related to the critical halo mass above which efficient gas cooling and compaction become possible \citep{Dekel_2006, Dekel_2009}. This is further supported by the Sérsic index trends shown in Appendix~\ref{app:sersic}, where a decrease in $n$ is observed at approximately the same stellar mass, consistent with a structural change from more irregular, disk-dominated morphologies at low masses to more bulge-dominated systems at high masses.

The characteristic crossover mass of $\log(M_*/M_\odot) \sim 9.5$ is broadly consistent with the mass scale identified in previous studies as the threshold above which star-forming galaxies develop significant bulge components and experience elevated dust attenuation (e.g. \citealt{Wuyts_2011, Lang_2014, Martis2016, Mosleh_2017}), as well as with the mass scale at which the star-forming main sequence \citep{Speagle2014} begins to show signs of central quenching and compaction in both observations and simulations \citep{Tacchella_2016, roper_2023}. 

\subsection{Inside-Out and Outside-In Formation at Low and High Masses}\label{sec:inside-out}
The multi-wavelength size--mass relation reveals two physically distinct regimes on either side of the crossover mass, which we interpret in terms of differing modes of mass assembly and star formation.

Above the crossover mass ($\log(M_*/M_\odot) \gtrsim 9.5$), galaxies are systematically smaller at rest-frame NIR wavelengths than in the rest-frame optical, and the size--mass slope is steeper at shorter wavelengths. This behaviour is consistent with inside-out growth, in which the central regions of galaxies assemble their stellar mass first and host an older, more evolved, and more compact stellar population, while star formation continues in the extended outer disk (e.g. \citealt{Mo_1998, vanDokkum_2010, Nelson_2016}; S. MacFarland et al. in prep). A consistent picture of inside-out growth is reported for spatially-resolved studies of comparable masses, $\log(M_*/M_\odot) \gtrsim 9.5$ (e.g. \citealt{Rowlands2018, Pessa2023}); \citet{Huang2018} also note that low-redshift massive galaxies tend to be more extended than low-mass galaxies, compatible with the interpretation of the NIR tracing older stellar populations; the NIR emission is therefore more centrally concentrated than the optical light, which receives contributions from the extended star-forming disk. High dust obscuration in the central regions of these galaxies, as evidenced by the elevated $A_V$ values at $\log(M_*/M_\odot) > 9.5$ (Appendix~\ref{app:dust}), further enhances the apparent optical size by attenuating the compact central component. Together, these effects produce the steep size--wavelength slope and the large ratio of optical to NIR size observed at high masses. 

Below the crossover mass, $\log(M_*/M_\odot) \lesssim 9.5$, the opposite behaviour is observed: NIR sizes are larger than or comparable to optical sizes, and the size--mass slope is flatter at longer wavelengths. This suggests a qualitatively different structural configuration, consistent with outside-in formation, in which star formation is concentrated in the central regions while the outskirts are dominated by an older, more diffuse stellar population (e.g. \citealt{Elmegreen_2005, Fudamoto_2022}). In this regime, the UV and optical light traces the compact, centrally concentrated star-forming regions, yielding a smaller measured size, while the NIR light traces the extended, older stellar envelope, yielding a larger measured size. This configuration may reflect the early stages of disk assembly in low-mass galaxies, where the gravitational potential is not yet deep enough to sustain extended star formation and gas is preferentially funnelled to the center. This is compatible with the characteristic stellar mass of $\sim10^{10}$ which marks the bimodality between the blue and red sequences at low redshifts \citep{Dekel_2006, Cattaneo_2006} and with previously reported evolutionary modes for low-mass systems and low-redshift systems, including dwarf galaxies \citep{zhang_intro, Cheng2020}. In investigating the local Tully-Fisher relation (TFR; \citealt{tully-fisher}) in star-forming galaxies, \citet{simons2015_tullyfisher} also present $\log(M_*/M_\odot) = 9.5$ as the transitionary ``mass of disk formation", above which nearly all local ($z <0.375$) galaxies display and form disks but below which a galaxy may or may not form a disk such that the TFR has significant scatter. This interpretation is also consistent with \citet{rosi_completeness} who present that the star-forming galaxy main sequence experiences an inflection point at $\log(M_*/M_\odot) \sim 9.5$ across redshifts, which may indicate a critical mass above which a stable and self-regulatory disk may assemble in star-forming galaxies. 

The transition between these two regimes at $\log(M_*/M_\odot) \sim 9.5$ may therefore mark the stellar mass at which galaxies shift from centrally concentrated, outside-in star formation to extended, inside-out growth. 

\subsection{Comparison to Simulations}\label{sec:sims}
The observed trends in the multi-wavelength size--mass relation provide a set of quantitative constraints that can be compared to predictions from hydrodynamical simulations of galaxy formation. In particular, the wavelength dependence of the size--mass slope, the crossover mass, and the ratio of NIR to optical sizes are all observables that encode information about the spatial distribution of stellar populations, dust, and star formation within galaxies, and thus provide a stringent test of the physical processes implemented in simulations. This trend is qualitatively consistent with the picture proposed by \citet{roper_2023} using the FLARES simulations (\citealt{Lovell_2020, Vijayan2021}), in which galaxies above a critical mass develop sufficiently deep potential wells to sustain the cooling and fragmentation of gas into compact, dense star-forming clumps in the central regions, driving the formation of a compact bulge component. Below this mass, the shallower potential well favours a more diffuse and centrally concentrated mode of star formation. Pre-JWST studies have also noted this behaviour, with \citet{Pan2015} utilizing low-redshift SDSS data to conclude that the flat colour gradients of galaxies in the local universe  experience ``outside-in" formation at $M_* < 10^{10}~M_\odot$. Furthermore, using the FIREbox simulations \citep{Feldmann_2023}, \citet{benavides2025} also suggest that a morphology transition regime exists for galaxies between $9 < \log(M_*/M_\odot) < 10$, above which extended stellar disks become common.

\section{Conclusions}\label{sec:summary}
We present a multi-wavelength morphological analysis of 4,140 star-forming galaxies at $0.6 < z \leq 4$ using JWST/NIRCam imaging from the CANUCS and JWST in Technicolor surveys, spanning rest-frame wavelengths from $\sim 0.3$ to $3.2\,\mu$m across 19 medium- and wide-band filters. We measure single-component Sérsic profiles for each galaxy in each filter using \galfit, and use these measurements to construct the rest-frame optical and NIR size--mass relations, characterize the size--wavelength relation as a function of stellar mass and redshift, and derive a generalized parametrization of galaxy size as a function of rest-frame wavelength, stellar mass, and redshift simultaneously. We concurrently present a public release of the morphological catalogue of structural parameters for 41,202 galaxies across the five CANUCS NIRCam flanking fields in up to 29 JWST+HST filters. Our main conclusions are as follows.

\begin{enumerate}
    \item \textbf{Rest-frame optical and NIR size--mass relations.} The rest-frame optical ($0.5\,\mu$m) size--mass relation for star-forming galaxies at $0.6 < z \leq 4$ is well described by a log-linear relation and evolves as $\log R_{\mathrm{eff, opt}} = (-0.71 \pm 0.16)\times\log_{10}(1+z) + (0.89 \pm 0.06)$ at a fixed stellar mass of $M_* = 5 \times 10^{10}\,M_\odot$, in good agreement with previous HST and JWST studies \citep{vanderWel_2014, Mowla_2019, Ward_2024, Allen2024}. The rest-frame NIR ($1.5\,\mu$m) size--mass relation, presented for $0.6 < z \leq 2$, is consistent within $1\sigma$ of \citet{Martorano_sizemass} at $\log(M_*/M_\odot) \gtrsim 9.2$, with a shallower slope at lower masses reflecting the behaviour of the low-mass galaxy population not captured in previous high-mass-limited studies.

    \item \textbf{Wavelength dependence of the size--mass slope.} The multi-wavelength size--mass relation (Figure~\ref{fig:size-mass FINAL}) displays a systematic gradient in slope as a function of rest-frame wavelength across all redshift bins: the NIR size--mass relation is consistently shallower than the rest-frame optical, with the difference most pronounced at high stellar masses. This trend is quantified by the Spearman correlation between stellar mass and the size--wavelength slope ($\rho_{M_*} = -0.47$), with negligible dependence on redshift ($\rho_z = -0.01$). We suggest this trend is driven by centrally concentrated dust attenuation consistent with radial colour gradients.

    \item \textbf{A characteristic crossover mass at $\log(M_*/M_\odot) \sim 9.5$.} We identify a characteristic crossover mass at $\log(\bar{M}_*/M_\odot) = 9.50 \pm 0.13 $, stable across all four redshift bins, at which the size--mass relations measured across all filters converge and the wavelength dependence of galaxy size is minimized. Above this mass, galaxies are on average $10.48\%$ smaller at rest-frame $1\,\mu$m than at rest-frame $0.5\,\mu$m, increasing to $26.74\%$ at $\log(M_*/M_\odot) > 10.5$, consistent with inside-out growth and centrally concentrated dust attenuation. Below the crossover mass, NIR sizes are larger than or comparable to optical sizes, consistent with outside-in formation in which star formation is centrally concentrated while older stellar populations dominate the outskirts.

    \item \textbf{A generalized size--mass--wavelength parametrization.} We derive a generalized parametrization of the effective radius as a continuous function of rest-frame wavelength, stellar mass, and photometric redshift over $0.25\mu m < \lambda_\mathrm{RF} < 2.0\mu m$ (Equation~\ref{eq:FINAL}), with wavelength-dependent coefficients fit by power-law and linear functions. This relation reproduces observed galaxy sizes with a residual scatter of $\sim 0.05$\,dex, and provides a practical tool for estimating galaxy sizes at arbitrary rest-frame wavelengths without requiring direct morphological measurements in each filter.
\end{enumerate}

Together, these results demonstrate that rest-frame wavelength is a critical parameter in characterizing galaxy structure, and that multi-wavelength morphological measurements with JWST are essential for disentangling the effects of stellar population gradients, dust attenuation, and structural evolution across cosmic time. Future work includes extending this analysis to higher redshifts ($z > 4$), lower stellar masses, and quiescent galaxies will provide further constraints on the physical processes driving the wavelength dependence of galaxy morphology.

\begin{acknowledgments}
This research was enabled by grants 18 JWST-GTO1, 23JWGO2A13, 23JWGO2B15, and  24JWGO2A04 from the Canadian Space Agency and Discovery Grant and Discovery Accelerator funding from the Natural Sciences and Engineering Research Council (NSERC) of Canada to MS. This research used the Canadian Advanced Network For Astronomy Research (CANFAR) platform operated in partnership by the Canadian Astronomy Data Centre and The Digital Research Alliance of Canada with support from the National Research Council of Canada, the Canadian Space Agency, CANARIE, and the Canada Foundation for Innovation. MB, JJ acknowledge support from the ERC Grant FIRSTLIGHT \#101053208, Slovenian national research agency ARIS through grants N1-0238 and P1-0188, and ESA PRODEX Experiment Arrangements No. 4000146646 and 4000149972. AM acknowledges support from the Yavin Family Fund. DM acknowledges generous support from the Leonard and Jane Holmes Bernstein Professorship in Evolutionary Science. Support for programs JWST-GO-03362, provided through a grant from the STScI under NASA contract NAS5-03127, is acknowledged. This work is based on observations made with the NASA/ESA/CSA James Webb Space Telescope. The data were obtained from the Mikulski Archive for Space Telescopes (MAST) at the Space Telescope Science Institute (STSci), which is operated by the Association of Universities for Research in Astronomy, Inc., under NASA contract NAS 5-03127 for JWST. JWST observations are associated with programs JWST-GTO-1208, -4527, and GO-3362.
\end{acknowledgments}

\section*{Data Availability} 
Data products presented in this paper are available through the CANUCS website\footnote{https://niriss.github.io/data.html} and as MAST High-Level Science Products\footnote{https://archive.stsci.edu/hlsp/canucs}. The CANUCS DOI is doi:\dataset[10.17909/ph4n-6n76]{http://dx.doi.org/10.17909/ph4n-6n76}. The JWST in Technicolor DOI is doi:\dataset[10.17909/cyh7-mm53]{http://dx.doi.org/10.17909/cyh7-mm53}. The data release HLSP DOI is doi:\dataset[10.17909/18nv-np70]{http://dx.doi.org/10.17909/18nv-np70}.

\facilities{JWST/NIRCam, HST(STIS)}

\software{\texttt{GALFIT} \citep{Peng_2002, Peng_2010}, \texttt{astropy} \citep{astropy}, \texttt{EAzY} \citep{Brammer2008-eazy}, \texttt{Dense Basis} \citep{Iyer_2017, Iyer_2019}, \texttt{photutils} \citep{bradley_photutils}}

\clearpage
\appendix
\onecolumngrid
\FloatBarrier
\section{Galaxy Parameter Distributions}\label{app:num sources}
\begin{figure}[htp!]
\centering
\includegraphics[width=0.72\linewidth]{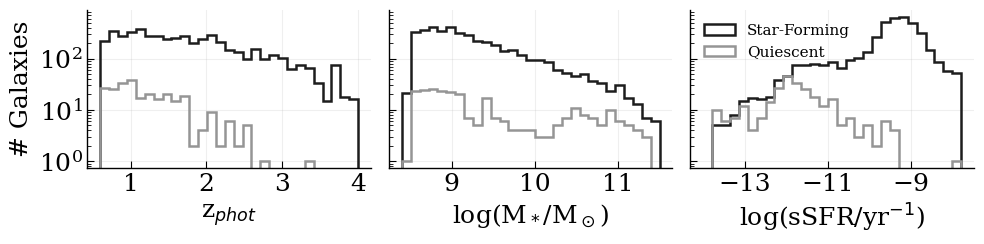}
\caption{Distributions of photometric redshift $z_\mathrm{phot}$ (left), stellar mass $\log(M_*/M_\odot)$ (centre), and specific star formation rate sSFR (right) for star-forming (blue) and quiescent (red) galaxies in the sample following the quality cuts and UVJ classification described in Section~\ref{sec:data selection}. }
\label{fig:UVJ-params}
\end{figure}

Figure~\ref{fig:UVJ-params} shows the distributions of photometric redshift, stellar mass, and specific star formation rate (sSFR) for the star-forming and quiescent subsamples identified via UVJ selection (Section~\ref{sec:data selection}). The star-forming sample spans $0.6 < z \leq 4$ and $8.5 < \log(M_*/M_\odot) \leq 11.5$, with the bulk of the population at $\log(M_*/M_\odot) \sim 9$--$10$. The quiescent sample is distributed to higher stellar masses and lower sSFR values, as expected given the UVJ selection criteria and the mass-completeness limits described in Section~\ref{sec:data selection}. The median stellar mass, redshift, and sSFR of the star-forming sample for each redshift bin are reported in Table \ref{tab:medians-bins}.

\begin{table}[htp!]
\centering
\begin{tabular}{lccc}
\hline\hline
Redshift & $\log(M_{*{med}}/M_\odot)$ & $z_{med}$ & $\log(sSFR_{med}/yr^{-1})$\\
\hline
$0.6 < z \leq 1.0$ & 8.68 & 0.81 & -9.74  \\
$1.0 < z \leq 1.5$ & 8.88 & 1.24 & -9.43  \\
$1.5 < z \leq 2.0$ & 9.00 & 1.75 & -9.29  \\
$2.0 < z \leq 4.0$ & 9.12 & 2.53 & -9.02  \\
\hline
\end{tabular}
\caption{Median values of stellar mass, redshift, and sSFR for the star-forming sample per redshift bin.}
\label{tab:medians-bins}
\end{table}

Table~\ref{tab:num-sources} lists the number of star-forming galaxies in each redshift and stellar mass bin used in the analysis. The sample is dominated by low-mass galaxies at $8.5 < \log(M_*/M_\odot) \leq 9.5$, which account for $\sim72\%$ of the total sample, reflecting both the steepness of the stellar mass function at low masses and the depth of the CANUCS-Technicolor imaging. The highest-mass bin ($10.5 < \log(M_*/M_\odot) \leq 11.5$) is the least populated, comprising only $\sim5\%$ of the sample. The most populated redshift bin is $2 < z \leq 4$, which spans the widest redshift interval and encompasses Cosmic Noon, the epoch of peak star formation activity \citep{Madau_Dickinson_2014}.

\begin{table}[htp!]
\centering
\begin{tabular}{lcccc}
\hline\hline
Redshift & \multicolumn{3}{c}{$\log(M_*/M_\odot)$} & Total \\
\cline{2-4}
 & $8.5$--$9.5$ & $9.5$--$10.5$ & $10.5$--$11.5$ & \\
\hline
$0.6 < z \leq 1.0$ & 435 & 126 & 30 & 591 \\
$1.0 < z \leq 1.5$ & 756 & 228 & 58 & 1,042 \\
$1.5 < z \leq 2.0$ & 688 & 194 & 66 & 948 \\
$2.0 < z \leq 4.0$ & 1,088 & 399 & 72 & 1,559 \\
\hline
Total & 2967 & 947 & 226 & 4,140 \\
\hline
\end{tabular}
\caption{Number of star-forming galaxies in each redshift and stellar mass bin comprising the final sample of 4,140 galaxies used in this analysis. Mass bins reflect the completeness limits described in Equation~\ref{eq:completeness}.}
\label{tab:num-sources}
\end{table}

\FloatBarrier
\section{\galfit\ Morphological Fitting}\label{app:galfit}
\begin{figure*}[ht!]
\centering
\includegraphics[width=\linewidth]{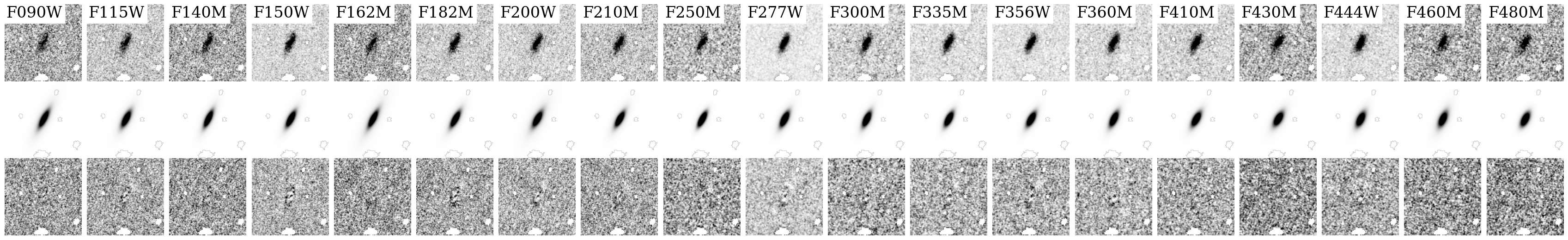}
\caption{Example of the single-component \galfit\ fitting output for galaxy CANUCS-A370-2200497 ($\log(M_*/M_\odot) = 9.02 \pm 0.121$, $z  = 1.99 \pm 0.101$) across nineteen JWST/NIRCam filters. }
\label{fig:five galfit}
\end{figure*}

Figure~\ref{fig:five galfit} shows an example of the single-component \galfit\ fitting output for a representative galaxy (CANUCS-A370-2200497; $\log(M_*/M_\odot) = 9.02 \pm 0.121$, $z  = 1.99 \pm 0.101$) across available JWST/NIRCam filters across the full wavelength range used in this analysis. For each filter, three panels are shown from top to bottom: the background-subtracted science image with neighbouring sources masked (in white), the best-fit single-component Sérsic model convolved with the filter-specific PSF, and the residual image obtained by subtracting the model from the science image. The residuals are consistent with noise across all filters, with no systematic over- or under-subtraction in the central regions, indicating that a single-component Sérsic profile provides an adequate description of the surface brightness distribution for this galaxy. 

\FloatBarrier

\section{Stellar Mass Completeness}\label{app:mass-completeness}
\begin{figure}[t!]
\centering
\includegraphics[width=0.6\linewidth]{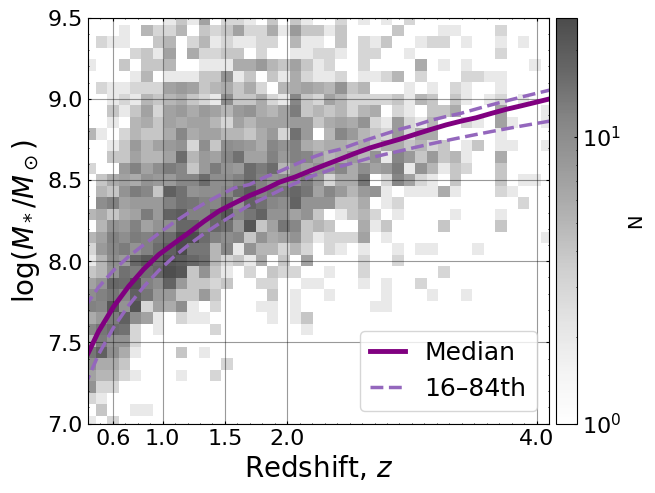}
\caption{Stellar mass as a function of photometric redshift for star-forming galaxies in the five CANUCS NIRCam flanking fields between $0.6 < z \leq 4$. The median (solid), 16th, and 84th percentile (dashed) stellar mass completeness limits estimated following \citet{rosi_completeness} are shown in purple, and completeness limits outlined in Equation~\ref{eq:completeness} are adopted.}
\label{fig:rosi-plot}
\end{figure}

Accurate characterization of the size--mass relation at low stellar masses requires a careful assessment of the mass completeness of the sample. We estimate the stellar mass completeness limit following the method of \citet{rosi_completeness}, which we briefly summarize here: mock spectral energy distributions (SEDs) of post-starburst galaxies are generated using \db\ across a grid of redshift and stellar mass bins, adopting a dust attenuation of $A_V = 0.3$\,mag as a representative value for the faint, low-mass galaxy population. Post-starburst galaxies are chosen as the reference population because they represent the faintest galaxies at a given stellar mass, and thus define a conservative lower bound on the detectable mass at each redshift. Each mock SED is compared to the observed galaxies in the photometric catalogue using a cosine similarity metric, and a selection function is constructed to estimate the completeness fraction as a function of redshift and stellar mass. The resulting mass completeness limits, defined as the stellar mass above which the sample is 50\% complete, are shown as a function of redshift in Figure~\ref{fig:rosi-plot} and are presented in Equation~\ref{eq:completeness}. Figure~\ref{fig:rosi-plot} shows the stellar masses of all star-forming galaxies in the CANUCS NCFs between $0.6 < z \leq 0.4$ as a function of photometric redshift, with the median, 16th, and 84th percentile completeness limits overplotted. The median completeness limit rises with redshift, as observed in Equation \ref{eq:completeness}, reflecting the decreasing depth of the survey in terms of stellar mass at higher redshifts. The adopted mass limits (Equation~\ref{eq:completeness}) are chosen conservatively to ensure that the sample remains complete across the full redshift range.

\section{Rest-Frame Wavelength Bins}\label{wav-bins}
Table~\ref{tab:wav bins} lists the number of individual galaxy-filter measurements in each rest-frame wavelength bin of width $0.15\,\mu$m used in the weighted least-squares regression described in Section~\ref{sec:size_mass_wav}. Bins are constructed by sorting all galaxy measurements by their rest-frame wavelength, computed from the filter effective wavelength and the photometric redshift of each galaxy. The sample peaks in number density around $0.4$--$0.7\,\mu$m, corresponding to the rest-frame optical regime most densely sampled by the JWST/NIRCam filter set at the median redshifts of our sample. Number counts decline steeply beyond $\sim 2\,\mu$m, as only the lowest-redshift galaxies contribute measurements at these rest-frame wavelengths via the longest-wavelength filters. The shaded rows indicate the bins retained for the WLS regression and functional parametrization of the size--mass--wavelength relation over $0.25\mu m < \lambda_\mathrm{RF} < 2.0\mu m$ (Equation~\ref{eq:FINAL}).

\begin{table}[ht!]
\centering
\small
\begin{tabular}{cc|cc}
\hline\hline
Wavelength Bin ($\mu$m) & Number of Measurements &
Wavelength Bin ($\mu$m) & Number of Measurements \\
\hline

$(0.099, 0.25]$ & 842 &
\cellcolor{band}$(1.60, 1.75]$ & 3874 \\

\cellcolor{band}$(0.25, 0.40]$ & 5769 &
\cellcolor{band}$(1.75, 1.90]$ & 2869 \\

\cellcolor{band}$(0.40, 0.55]$ & 9366 &
\cellcolor{band}$(1.90, 2.05]$ & 2481 \\

\cellcolor{band}$(0.55, 0.70]$ & 9148 &
$(2.05, 2.20]$ & 1703 \\

\cellcolor{band}$(0.70, 0.85]$ & 7955 &
$(2.20, 2.35]$ & 1307 \\

\cellcolor{band}$(0.85, 1.00]$ & 7709 &
$(2.35, 2.50]$ & 897 \\

\cellcolor{band}$(1.00, 1.15]$ & 7061 &
$(2.50, 2.65]$ & 656 \\

\cellcolor{band}$(1.15, 1.30]$ & 6178 &
$(2.65, 2.80]$ & 340 \\

\cellcolor{band}$(1.30, 1.45]$ & 5520 &
$(2.80, 2.95]$ & 134 \\

\cellcolor{band}$(1.45, 1.60]$ & 4813 &
$(2.95, 3.10]$ & 38 \\

\hline
\end{tabular}
\caption{Number of individual galaxy-filter measurements per rest-frame wavelength bin of width $0.15\,\mu$m. Shaded rows indicate the bins over $0.25\mu m < \lambda_\mathrm{RF} < 2.0\mu m$ retained for the weighted least-squares regression and functional parametrization of the size--mass--wavelength relation (Section~\ref{sec:size_mass_wav} and Equation~\ref{eq:FINAL}). Unshaded rows that fall outside this range are excluded from the functional fits due to limited sample size ($< 2,000$ galaxies).}
\label{tab:wav bins}
\end{table}

\FloatBarrier

\section{Catalogue Description}\label{app:cat-params}

\begin{deluxetable}{lcl}
\tablenum{6}
\tablecaption{Description of columns in the CANUCS-Technicolor NCF Morphology Catalogue. 
\textcolor{purple}{S+25} represents the parent photometric sample given by \citealt{Sarrouh2025}.}
\label{tab:struct_params}
\tablewidth{0pt}
\tabletypesize{20}
\tablehead{
\colhead{Column Name} & \colhead{Unit} & \colhead{Description}
}
\startdata
\multicolumn{3}{l}{\textit{Identification and Redshift}} \\
\hline
\texttt{SOURCE} & --- & Source identifier from \textcolor{purple}{S+25} \\
\texttt{Z\_ML} & --- & Maximum-likelihood photometric redshift from \eazy, used for physical size conversions \\
\texttt{Z\_ML\_CHI2} & --- & Chi-squared value at the maximum-likelihood photometric redshift \\
\texttt{RA} & deg & Right ascension (J2000) from \textcolor{purple}{S+25} \\
\texttt{DEC} & deg & Declination (J2000) from \textcolor{purple}{S+25} \\
\hline
\multicolumn{3}{l}{\textit{Median, Rest-Frame Optical, and Rest-Frame Near-Infrared Size}} \\
\hline
\texttt{R\_KPC\_MEDIAN} & kpc & Median effective radius across all filters in which a valid fit is available \\
\texttt{R\_KPC\_MEDIAN\_ERR} & kpc & Uncertainty on the median effective radius \\
\texttt{R\_KPC\_REST\_OPT} & kpc & Rest-frame $0.5\,\mu$m effective radius from the filter with $\lambda_{EFF}$ closest to $0.5\,\mu$m \\
\texttt{R\_KPC\_REST\_OPT\_ERR} & kpc & Uncertainty on the rest-frame $0.5\,\mu$m effective radius \\
\texttt{R\_REST\_OPT\_FILT} & --- & Filter corresponding to the rest-frame $0.5\,\mu$m \\
\texttt{R\_KPC\_REST\_NIR} & kpc & Rest-frame $1.5\,\mu$m effective radius from the filter with $\lambda_{EFF}$ closest to $1.5\,\mu$m ($z \leq 2$) \\
\texttt{R\_KPC\_REST\_NIR\_ERR} & kpc & Uncertainty on the rest-frame $1.5\,\mu$m effective radius \\
\texttt{R\_REST\_NIR\_FILT} & --- & Filter corresponding to the rest-frame $1.5\,\mu$m \\
\hline
\multicolumn{3}{l}{\textit{Quality Flags}} \\
\hline
\texttt{USE\_PA} & --- & (1) if reliable in $\geq 10$ bands; (0) if unreliable: \texttt{PA\_ERR\_\{BAND\}} $> 90^\circ$ and \texttt{Q\_\{BAND\}} $< 0.7$ \\
\texttt{USE\_N} & --- & (1) if $0.1 < n < 10$ in $\geq 10$ bands; (0) otherwise \\
\texttt{USE\_Q} & --- & (1) if $0.05 < Q < 1$ in $\geq 10$ bands; (0) otherwise \\
\texttt{USE\_FLAG} & --- & Combined $n$ and $Q$ flag: (1) if both converge within bounds in $\geq 10$ bands; (0) otherwise \\
\hline
\multicolumn{3}{l}{\textit{Derived Physical Sizes per Photometric Band}} \\
\hline
\texttt{R\_ARCSEC\_\{BAND\}} & arcsec & Best-fit effective (half-light) radius along the semi-major axis \\
\texttt{R\_ARCSEC\_ERR\_\{BAND\}} & arcsec & Uncertainty on the effective radius from \galfit \\
\texttt{R\_KPC\_\{BAND\}} & kpc & Effective radius along the semi-major axis, converted to physical kiloparsecs using $z_\mathrm{ml}$ \\
\texttt{R\_KPC\_ERR\_\{BAND\}} & kpc & Uncertainty on the effective radius; does not include the contribution from $z_\mathrm{ml}$ uncertainty \\
\texttt{R\_CIRC\_KPC\_\{BAND\}} & kpc & Circularized effective radius, defined as $R_\mathrm{circ} = R_\mathrm{eff}\sqrt{Q}$ \\
\texttt{R\_CIRC\_KPC\_ERR\_\{BAND\}} & kpc & Uncertainty on the circularized effective radius \\
\hline
\multicolumn{3}{l}{\textit{Fitted Structural Parameters per Photometric Band}} \\
\hline
\texttt{M\_\{BAND\}} & mag & Best-fit total apparent AB magnitude from \galfit \\
\texttt{M\_ERR\_\{BAND\}} & mag & Uncertainty in the fitted AB magnitude \\
\texttt{N\_\{BAND\}} & --- & Best-fit Sérsic index \\
\texttt{N\_ERR\_\{BAND\}} & --- & Uncertainty in the Sérsic index \\
\texttt{Q\_\{BAND\}} & --- & Best-fit axis ratio ($b/a$; minor to major axis) \\
\texttt{Q\_ERR\_\{BAND\}} & --- & Uncertainty in the axis ratio \\
\texttt{PA\_\{BAND\}} & deg & Best-fit position angle, measured east of north \\
\texttt{PA\_ERR\_\{BAND\}} & deg & Uncertainty in the position angle \\
\texttt{X\_CEN\_\{BAND\}} & pix & $X$-centroid of the best-fit Sérsic profile in the filter cutout \\
\texttt{X\_CEN\_ERR\_\{BAND\}} & pix & Uncertainty in the $X$-centroid position \\
\texttt{Y\_CEN\_\{BAND\}} & pix & $Y$-centroid of the best-fit Sérsic profile in the filter cutout \\
\texttt{Y\_CEN\_ERR\_\{BAND\}} & pix & Uncertainty in the $Y$-centroid position \\
\texttt{RA\_GAL\_\{BAND\}} & deg & Right ascension converted from the $X$-centroid \\
\texttt{DEC\_GAL\_\{BAND\}} & deg & Declination converted from the $Y$-centroid \\
\texttt{SNR\_\{BAND\}} & --- & Photometric signal-to-noise ratio from \textcolor{purple}{S+25} \\
\texttt{GAL\_ERR\_\{BAND\}} & --- & Morphological uncertainty parameter, defined as $R_\mathrm{eff}/\sigma_{R_\mathrm{eff}}$ (see Section~\ref{subsec:galfit}) \\
\enddata
\end{deluxetable}
\onecolumngrid

The CANUCS-Technicolor NCF Morphology Catalogue is released alongside this paper and contains single-component Sérsic fit parameters measured with \galfit\ for $\sim$41,000 galaxies across the five CANUCS NIRCam flanking fields in up to 19 JWST/NIRCam filters, as described in Section~\ref{sec:data products}. For each galaxy, structural parameters are provided for every filter in which a fit was performed, with column names of the form \texttt{\{PARAM\}\_\{BAND\}}, where \texttt{\{BAND\}} denotes the filter name (e.g., \texttt{F277W}). Derived physical sizes in kiloparsecs are computed using the maximum-likelihood photometric redshift $z_\mathrm{ml}$ from \eazy. All sizes are PSF-deconvolved. Uncertainties are propagated directly from the \galfit\ covariance matrix and do not include contributions from photometric redshift uncertainty. A quality flag (\texttt{USE\_FLAG}) is provided for each source indicating whether the fitted Sérsic index and axis ratio fall within the imposed \galfit\ bounds; users are strongly encouraged to apply this flag when using the catalogue for scientific analysis. A complete description of all catalogue columns is given in Table~\ref{tab:struct_params}.

\section{Weighted Least-Squares Coefficients by Wavelength Bin}\label{app:wls-coeffs}
Table~\ref{tab:filter_coefficients} lists the best-fit coefficients of the weighted least-squares regression described in Section~\ref{sec:size_mass_wav} for each rest-frame wavelength bin over $0.25\mu m < \lambda_\mathrm{RF} < 2.0\mu m$. In each bin, the relation takes the form,
\begin{align*}
\log\left(\frac{R_\mathrm{eff}}{\mathrm{kpc}}\right) = m(\lambda)\,\log(M_*/M_\odot) + z(\lambda)\,z_\mathrm{phot} + K(\lambda),
\end{align*}
where $m(\lambda)$ is the stellar mass coefficient, $z(\lambda)$ is the redshift coefficient, and $K(\lambda)$ is the constant intercept term. Coefficients and their $1\sigma$ uncertainties are derived from the weighted least-squares fit in each bin independently, with each galaxy weighted by its \galfit\ morphological uncertainty (Section~\ref{subsec:galfit}). The systematic decrease in $m(\lambda)$ with increasing wavelength reflects the flattening of the size--mass slope towards the rest-frame NIR, as discussed in Section~\ref{sec:size_mass_wav}. The redshift coefficient $z(\lambda)$ is consistent with a constant value across all bins, motivating the linear parametrization adopted in Equation~\ref{eq:FINAL}. The functional forms fit to these coefficients over the full wavelength range are shown in Figure~\ref{fig:coeffs_fits}.

\begin{table*}[ht!]
\centering
\begin{tabular}{cccc}
\hline\hline
$\lambda_\mathrm{RF}$ ($\mu$m) & $m(\lambda)$ & $z(\lambda)$ & $K(\lambda)$ \\
\hline
$(0.25, 0.40]$ & $0.2623 \pm 0.0044$ & $-0.1087 \pm 0.0043$ & $-2.0079 \pm 0.0458$ \\
$(0.40, 0.55]$ & $0.2530 \pm 0.0031$ & $-0.1137 \pm 0.0030$ & $-1.8956 \pm 0.0317$ \\
$(0.55, 0.70]$ & $0.2223 \pm 0.0031$ & $-0.0950 \pm 0.0030$ & $-1.6364 \pm 0.0319$ \\
$(0.70, 0.85]$ & $0.1969 \pm 0.0033$ & $-0.1016 \pm 0.0030$ & $-1.3838 \pm 0.0340$ \\
$(0.85, 1.00]$ & $0.1810 \pm 0.0033$ & $-0.0872 \pm 0.0032$ & $-1.2630 \pm 0.0341$ \\
$(1.00, 1.15]$ & $0.1676 \pm 0.0034$ & $-0.1004 \pm 0.0037$ & $-1.1154 \pm 0.0351$ \\
$(1.15, 1.30]$ & $0.1488 \pm 0.0035$ & $-0.0836 \pm 0.0047$ & $-0.9616 \pm 0.0358$ \\
$(1.30, 1.45]$ & $0.1581 \pm 0.0041$ & $-0.0882 \pm 0.0062$ & $-1.0310 \pm 0.0407$ \\
$(1.45, 1.60]$ & $0.1504 \pm 0.0042$ & $-0.1444 \pm 0.0075$ & $-0.8954 \pm 0.0420$ \\
$(1.60, 1.75]$ & $0.1207 \pm 0.0049$ & $-0.1162 \pm 0.0101$ & $-0.6307 \pm 0.0493$ \\
$(1.75, 1.90]$ & $0.1596 \pm 0.0058$ & $-0.1041 \pm 0.0153$ & $-1.0375 \pm 0.0576$ \\
\hline
\end{tabular}
\caption{Best-fit coefficients and $1\sigma$ uncertainties of the weighted least-squares regression in each rest-frame wavelength bin over $0.25\mu m < \lambda_\mathrm{RF} < 2.0\mu m$. The relation in each bin is $\log(R_\mathrm{eff}/\mathrm{kpc}) = m(\lambda)\,\log(M_*/M_\odot) + z(\lambda)\,z_\mathrm{phot} + K(\lambda)$, where $m(\lambda)$, $z(\lambda)$, and $K(\lambda)$ are the stellar mass, redshift, and constant coefficients respectively. The systematic decrease in $m(\lambda)$ with increasing wavelength reflects the flattening of the size--mass slope towards the rest-frame NIR. The full wavelength-dependent parametrization of these coefficients is given in Equation~\ref{eq:FINAL} and Figure~\ref{fig:coeffs_fits}.}
\label{tab:filter_coefficients}
\end{table*}

\FloatBarrier

\section{Residuals of the Size--Mass--Wavelength Relation}\label{app:residuals}
\begin{figure}[t]
\centering
\includegraphics[width=0.6\linewidth]{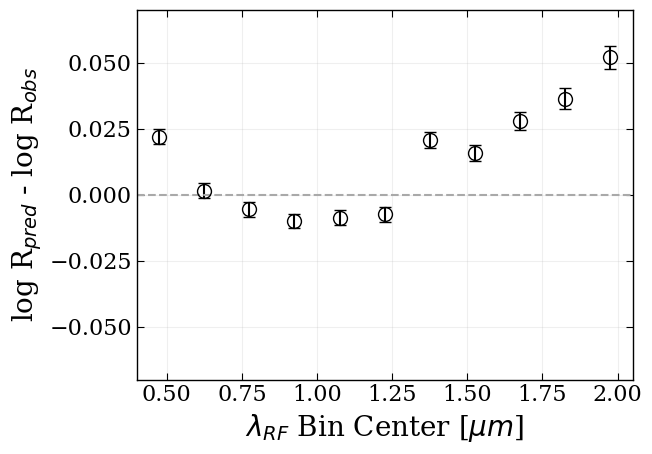}
\caption{Residuals $\Delta \log R_\mathrm{eff} = \log R_\mathrm{eff,\,obs} - \log R_\mathrm{eff,\,pred}$ between observed and predicted effective radii from Equation~\ref{eq:FINAL} as a function of rest-frame wavelength. Points show the weighted mean residual in each wavelength bin and error bars indicate the weighted standard deviation. The dashed line at $\Delta \log R_\mathrm{eff} = 0$ indicates perfect agreement. The intrinsic scatter of $\pm 0.05$\,dex and mean offset of $\mu_\Delta = 0.01$\,dex confirm that the relation provides an accurate and nearly unbiased prediction of galaxy size across the fitted wavelength range.}
\label{fig:residuals}
\end{figure}

Figure~\ref{fig:residuals} shows the residuals between the observed effective radii and those predicted by the generalized size--mass--wavelength relation (Equation~\ref{eq:FINAL}), computed in each rest-frame wavelength bin over $0.25\mu m < \lambda_\mathrm{RF} < 2.0\mu m$. The residuals are defined as $\Delta \log R_\mathrm{eff} = \log R_\mathrm{eff,\,obs} - \log R_\mathrm{eff,\,pred}$, with uncertainties given by the weighted standard deviation within each bin. The relation performs with an intrinsic scatter of $\pm 0.05$\,dex and a mean residual of $\mu_\Delta = 0.01$\,dex, indicating a slight systematic over-prediction of galaxy size. The residuals exhibit a mild but systematic variation with wavelength with a dip at mid-range wavelengths ($\sim 0.75$--$1.25\,\mu$m), suggesting that the power-law functional forms adopted for $m(\lambda)$ and $K(\lambda)$ (Equation~\ref{eq:FINAL}) provide a reasonable but not perfect description of the wavelength dependence of galaxy size; a more flexible parametrization may better capture the detailed shape of the coefficients across the full wavelength range. The amplitude of these systematic residuals remains small ($\lesssim 0.05$\,dex), and we defer a more detailed investigation of higher-order wavelength dependencies to future work. The small mean offset and absence of wavelength-dependent trends together demonstrate that Equation~\ref{eq:FINAL} provides a compact and accurate parametrization of galaxy size as a joint function of rest-frame wavelength, stellar mass, and redshift.

\FloatBarrier
\section{Sérsic Index as a Function of Stellar Mass}\label{app:sersic}
\begin{figure}[htp!]
\centering
\includegraphics[width=0.9\linewidth]{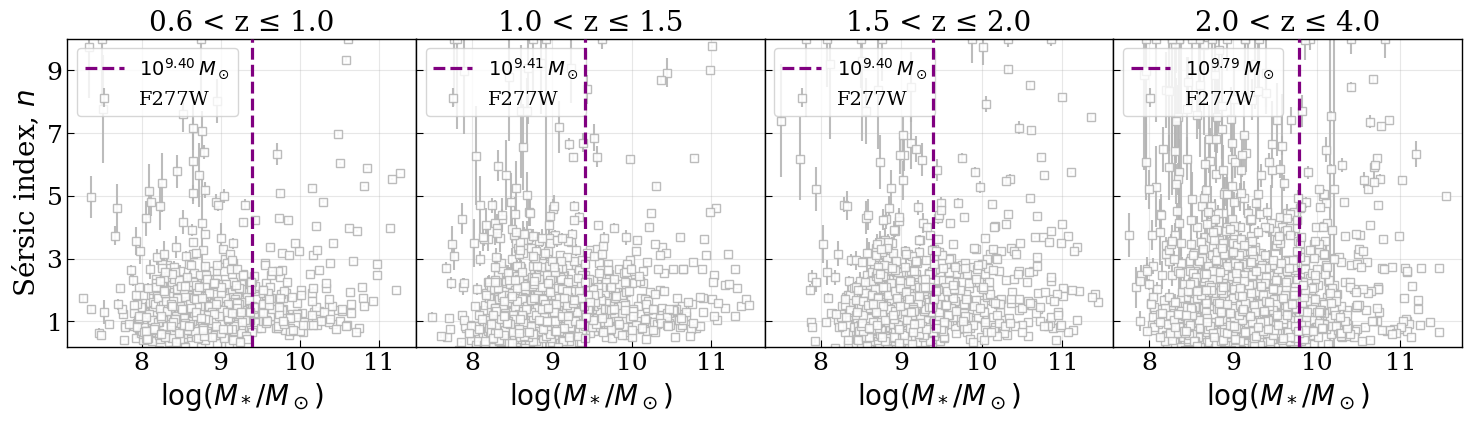}
\caption{Sérsic index $n$ as a function of stellar mass for star-forming galaxies in each redshift bin, measured in JWST/NIRCam F277W. The vertical dashed purple line indicates the crossover mass in each redshift bin as derived from the multi-wavelength size--mass relation (Figure~\ref{fig:size-mass FINAL}). The transition in Sérsic index at the crossover mass is consistent with a structural change from irregular or disc-dominated morphologies at low masses to more bulge-dominated systems at high masses.}
\label{fig:n v m}
\end{figure}

Figure~\ref{fig:n v m} shows the Sérsic index $n$ as a function of stellar mass measured in F277W for star-forming galaxies per redshift bin; we utilize F277W as it is the deepest filter available in the CANUCS-Technicolor dataset and provides the highest $S/N$ morphological measurements, minimizing the contribution of fitting noise to the observed trends. The crossover mass identified in Section~\ref{sec:multi-wav-sizemass} is overplotted in each panel as a vertical dashed line.

Two trends are apparent. First, the median Sérsic index increases with stellar mass, with $n_{med, ~8.5-9.5} \sim 1.4$ and $n_{med, ~10.5-11.5} \sim 1.7$; second, the uncertainty on $n$ is significantly larger ($\sim 2-2.5\times$ larger) below the crossover mass ($\log(M_*/M_\odot) \lesssim 9.5$) than at high masses across redshift bins, reflecting the lower signal-to-noise ratio of the morphological fits in this regime. Notably, this provides independent structural evidence that $\log(M_*/M_\odot) \sim 9.5$ represents a physically meaningful transition scale in galaxy morphology. 

\FloatBarrier
\section{Dust Attenuation}\label{app:dust}
Table \ref{tab:Av_mass_redshift} presents median values of $A_V$ as a function of both mass and redshift, and quantifies the sharp increase $A_V$ as a function of stellar mass. The sharp rise in $A_V$ with the crossover mass further suggests that the transition at $\log(M_*/M_\odot) \sim 9$--$10$ is not merely a morphological change but reflects a physical transition in the dust content and interstellar medium properties of star-forming galaxies at this mass scale. This is consistent with the picture in which galaxies above the crossover mass have sufficiently deep potential wells to retain metal-enriched gas and sustain higher dust-to-gas ratios, while lower-mass galaxies lose a significant fraction of their metal-enriched gas through stellar feedback \citep{mass-metallicity}, and that galaxy mass is the primary driver of dust attenuation \citep{Lorenz2023}.

\begin{table}[h!]
\caption{Median $A_V$ values per stellar mass and redshift bin for star-forming galaxies.}
\centering
\begin{tabular}{lccc}
\hline\hline
Redshift & \multicolumn{3}{c}{$\log(M_*/M_\odot)$} \\
\cline{2-4}
 & $8$--$9$ & $9$--$10$ & $10$--$11.5$\\
\hline
$0.6 < z \leq 1.0$ & 0.027 & 0.338 & 0.868 \\
$1.0 < z \leq 1.5$ & 0.084 & 0.667 & 1.827 \\
$1.5 < z \leq 2.0$ & 0.153 & 0.830 & 1.949 \\
$2.0 < z \leq 4.0$ & 0.252 & 0.789 & 2.540 \\
\hline
\end{tabular}
\label{tab:Av_mass_redshift}
\end{table}

\begin{figure}[hp!]
\centering
\includegraphics[width=0.79\linewidth]{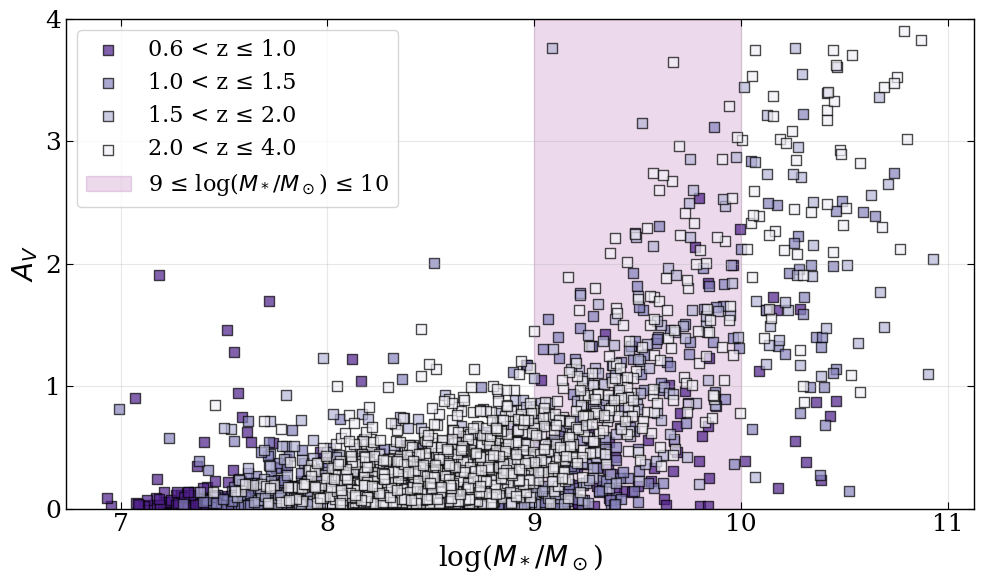}
\caption{$V$-band dust attenuation $A_V$ as a function of stellar mass for star-forming galaxies in the four redshift bins ($0.6 < z \leq 1.0$, $1.0 < z \leq 1.5$, $1.5 < z \leq 2.0$, and $2.0 < z \leq 4.0$), shown in increasingly lighter shades of grey. $A_V$ values are derived from \db\ SED fitting. The dashed vertical lines at $\log(M_*/M_\odot) = 9$ and $10$ bracket the mass range over which $A_V$ increases sharply, coinciding with the crossover mass identified in the multi-wavelength size--mass relation (Figure~\ref{fig:size-mass FINAL}). The sharp increase in dust attenuation in this regime is consistent with centrally concentrated dust being a primary driver of the wavelength-dependent size differences observed at high stellar masses.}
\label{fig:Av vs mass}
\end{figure}

\clearpage
\bibliography{bib}{}
\bibliographystyle{aasjournalv7}

\end{document}